\documentclass[12pt, onecolumn, draftclsnofoot]{IEEEtran}
\usepackage{amsfonts}
\usepackage{times}
\usepackage{graphicx}
\usepackage{epstopdf}
\usepackage{latexsym}
\usepackage{dsfont}
\usepackage{amssymb}
\usepackage{amsmath}
\usepackage{cite}
\usepackage{verbatim}
\usepackage{subfigure}

\def\bb0{{\mathbb{0}}}

\def\ba{{\mathbf{a}}}
\def\bb{{\mathbf{b}}}
\def\bc{{\mathbf{c}}}

\def\bff{{\mathbf{f}}}

\def\bh{{\mathbf{h}}}

\def\bp{{\mathbf{p}}}
\def\bq{{\mathbf{q}}}

\def\bt{{\mathbf{t}}}

\def\bv{{\mathbf{v}}}

\def\b0{{\mathbf{0}}}

\def\bH{{\mathbf{H}}}

\def\bQ{{\mathbf{Q}}}

\def\bV{{\mathbf{V}}}

\def\bX{{\mathbf{X}}}

\def\bbC{{\mathbb{C}}}

\def\bbE{{\mathbb{E}}}

\def\bbR{{\mathbb{R}}}

\def\bbZ{{\mathbb{Z}}}

\def\cN{\mathcal{N}}
\def\cO{\mathcal{O}}
\def\cP{\mathcal{P}}
\def\cQ{\mathcal{Q}}

\def\sf0{{\mathsf{0}}}

\usepackage{graphicx}
\usepackage{epstopdf}
\usepackage{enumerate}
\usepackage{algorithm}
\usepackage{amsmath}
\usepackage{mathrsfs}
\usepackage{float}
\usepackage{color}
\usepackage{makeidx}
\usepackage{bm}
\usepackage{cleveref}
\usepackage{url}
\usepackage{steinmetz}
\usepackage{varwidth}
\usepackage{soul}
\usepackage{bbm}

\newcommand{\sref}[1]{{Section}~\ref{#1}}

\def \rm {\mathrm}

\usepackage{amsfonts}
\usepackage{booktabs}
\usepackage{siunitx}
\usepackage{empheq}

\begin{document}
	\title{Sensing Aided Reconfigurable Intelligent Surfaces for  3GPP  5G Transparent Operation}
	
	\author{Shuaifeng Jiang, Ahmed Hindy, and Ahmed Alkhateeb \thanks{Shuaifeng Jiang and Ahmed Alkhateeb are with the School of Electrical, Computer, and Energy Engineering at Arizona State University - Email: \{s.jiang, alkhateeb\}@asu.edu. Ahmed Hindy is with Motorola Mobility LLC (a Lenovo company) - Email: ahmedhindy@motorola.com. This work was supported by the National Science Foundation (NSF) under Grant No. 2048021.} }
	
	\maketitle
	
	\begin{abstract}
		Can reconfigurable intelligent surfaces (RISs) operate in a standalone mode that is completely transparent to the 3GPP 5G initial access process? Realizing that may greatly simplify the deployment and operation of these surfaces and reduce the infrastructure control overhead. This paper investigates the feasibility of building standalone/transparent RIS systems and shows that one key challenge lies in determining the user equipment (UE)-side RIS beam reflection direction. To address this challenge, we propose to equip the RISs with multi-modal sensing capabilities (e.g., using wireless and visual sensors) that enable them to develop some perception of the surrounding environment and the mobile users. Based on that, we develop a machine learning framework that leverages the wireless and visual sensors at the RIS to select the optimal beams between the base station (BS) and users and enable 5G standalone/transparent RIS operation. Using a high-fidelity synthetic dataset with co-existing wireless and visual data, we extensively evaluate the performance of the proposed framework. Experimental results demonstrate that the proposed approach can accurately predict the BS and UE-side candidate beams, and that the standalone RIS beam selection solution is capable of realizing near-optimal achievable rates with significantly reduced beam training overhead.
	\end{abstract}


	\section{Introduction}
	Reconfigurable intelligent surfaces (RISs) have the potential of extending the coverage and reliability of millimeter wave (mmWave) and terahertz communication networks in 5G and beyond \cite{Basar_RIS,Pan21,Bjornson_RIS,Trichopoulos22,Yuanwei2021,ElMossallamy}. 
	In particular, these surfaces employ large numbers of reflecting elements which, when properly controlled, can reflect and focus the incident signals towards the wireless receiver. This enables the network to bypass blockages and maintain reliable link connections. 
	Figuring out the right configuration of these reflecting elements, however, is a challenging task that requires sufficient channel knowledge (for RIS-basestation and RIS-user sides) and large beam training and feedback overhead \cite{taha2021enabling}. 
	Prior work has mainly assumed that the RIS is controlled by the basestation/infrastructure which, while simplifying its operation, requires high channel estimation and control signaling overhead, which need to be captured in future releases of the 5G standard to ensure interoperability across different UE and network vendors.
	In this paper, we address the following question: \textbf{Can we build  \textit{standalone RISs} of which the operation is transparent to the 3GPP 5G protocol?} In particular, can the RIS assist the base station-user link without any coordination or feedback from any of them? We present the key challenges in realizing this standalone/transparent RIS operation vision and show how these challenges can be relaxed by employing sensing at the RIS surfaces. 
	
	\subsection{Prior Work}
	Prior work has extensively investigated the various aspects of the RIS system operation, including its channel estimation and beamforming \cite{mishra2019channel,yang2020intelligent,jensen2020optimal,taha2021enabling} and the application of machine learning (ML) to enhance RIS-aided communication systems \cite{taha2021enabling,Taha2020RL}. 
	For example, in \cite{mishra2019channel}, a channel estimation procedure for the RIS-aided communication systems was proposed relying on activating the RIS elements one by one, and estimating the end-to-end basestation-RIS-user channel at the basestation. To reduce the estimation overhead, \cite{yang2020intelligent} proposed to divide the reflecting elements into groups and estimate the effective channel for each group.  Alternatively, \cite{jensen2020optimal} developed a channel estimation approach that relies on varying the RIS phase shift configurations with each pilot symbol.  
	These channel estimation procedures in \cite{mishra2019channel, yang2020intelligent, jensen2020optimal}, however, can only be used if the RIS is controlled by the basestation and do not support standalone RIS operation because: (i) These approaches implicitly assume that the RIS reflection configuration is synchronized with the pilot transmissions, which requires the RIS coordination with the user and infrastructure and (ii) after channel estimation, the BS needs to feedback the estimated channel or the beamforming configuration to the RIS, which also needs dedicated signaling.

	Towards standalone RIS operation, \cite{taha2021enabling,Taha2020RL} developed what is known as the semi-passive RIS architecture which uses sparse active antenna elements to estimate the channels of the incident signals/pilots. Considering the 3GPP 5G initial access process \cite{3GPP300}, this semi-passive RIS architecture can enable the RIS to estimate the basestation-side channel leveraging the basestation synchronization signaling. However, since the users do not transmit any pilot signals before they receive the basestation synchronization signals, this semi-passive RIS architecture alone is not sufficient to enable full standalone RIS operation. In another context, integrating multi-modal sensing at the infrastructure and mobile users to aid the wireless communication decisions has been recently attracting interest for different use cases \cite{Wang18,Arvinte2019,Morais22,Rezaie2022,Alrabeiah2020a,Charan2022,Demirhan2022,Jiang_LiDAR,Demirhan_mgazine_radar}. For example, in \cite{Wang18,Arvinte2019,Rezaie2022}, the authors showed that position and orientation data could enable the mobile users to predict their optimal beam directions and reduce the beam search overhead. Further, in \cite{Alrabeiah2020a,Charan2022}, the visual data collected by cameras installed at the basestation was utilized to narrow down the beam search space. Similarly, \cite{Demirhan2022,Jiang_LiDAR} built real-world proof-of-concept prototypes that demonstrated the feasibility of using LiDAR and radar information to aid the mmWave beam selection process and significantly reduce the beam training overhead.

	
	\subsection{Contribution}
	In this paper, we propose a sensing-aided RIS operation that is transparent to the 3GPP 5G initial access process. In particular, we propose to equip the RIS with wireless and visual sensing that enable it to develop some perception of the surrounding communication environment and help it make efficient beam selection decisions. The main contributions of this paper can be summarized as follows:  
	\begin{itemize}
		\item We develop a decoupled BS-side and UE-side RIS beam design formulation that has low complexity yet can lead to near-optimal performance in realistic propagation scenarios. This decoupling is also essential for the proposed RIS standalone/transparent operation. 
		\item We introduce a sensing based perception framework for enabling 3GPP 5G standalone RIS operation. With the sensing capabilities, the RIS builds awareness about its surrounding environment, which can be utilized to guide the standalone RIS beam selection process. As one example, we show how the RIS can leverage visual sensors (cameras) to aid the selection of the UE-side beams. 
		\item We develop a machine learning framework that predicts the UE-side candidate beam sets based on the visual information. The proposed learning framework first detects the candidate UEs using an object detector. Then, a neural network (NN) architecture is proposed to predict the corresponding UE-side candidate beam set.
		\item We build a high-fidelity synthetic dataset that gathers co-existing wireless and visual data for the adopted RIS system. Using this large-scale dataset, we extensively evaluate the performance of the proposed sensing-aided RIS beam selection algorithms.  
	\end{itemize}
	\par
	Simulation results demonstrate the efficiency of the proposed algorithms in reducing the beam training overhead, achieving near-optimal data rates, and enabling standalone RIS operation that is compliant with the 3GPP 5G initial access process, highlighting a promising framework for future RIS-aided wireless communication systems. 
	
	\par
	The rest of the paper is organized as follows. Section \ref{System and Channel Models} explains the system and channel models. Section \ref{5G 3GPP Initial Access} introduces the outline of the 5G NR initial access procedure. Section \ref{5G 3GPP Standalone RIS} presents the key challenge of transparent 3GPP RIS. In Section \ref{Sensing for Standalone RIS Operation}, we demonstrate the idea of obtaining environment-awareness with sensing based perception, and present the ML framework and our standalone RIS beam selection. Section \ref{Deep Learning Modeling} explains the deep learning (DL) models for the ML framework. Simulation setup and results are presented in Section \ref{setup} and Section \ref{Simulation}. In \sref{Future Work}, we discuss the enabling features to extend the standalone RIS operation to more complex scenarios. Section \ref{Conclusion} concludes the paper.

	\begin{figure}[t]
		\centering
		\includegraphics[width=.75 \linewidth]{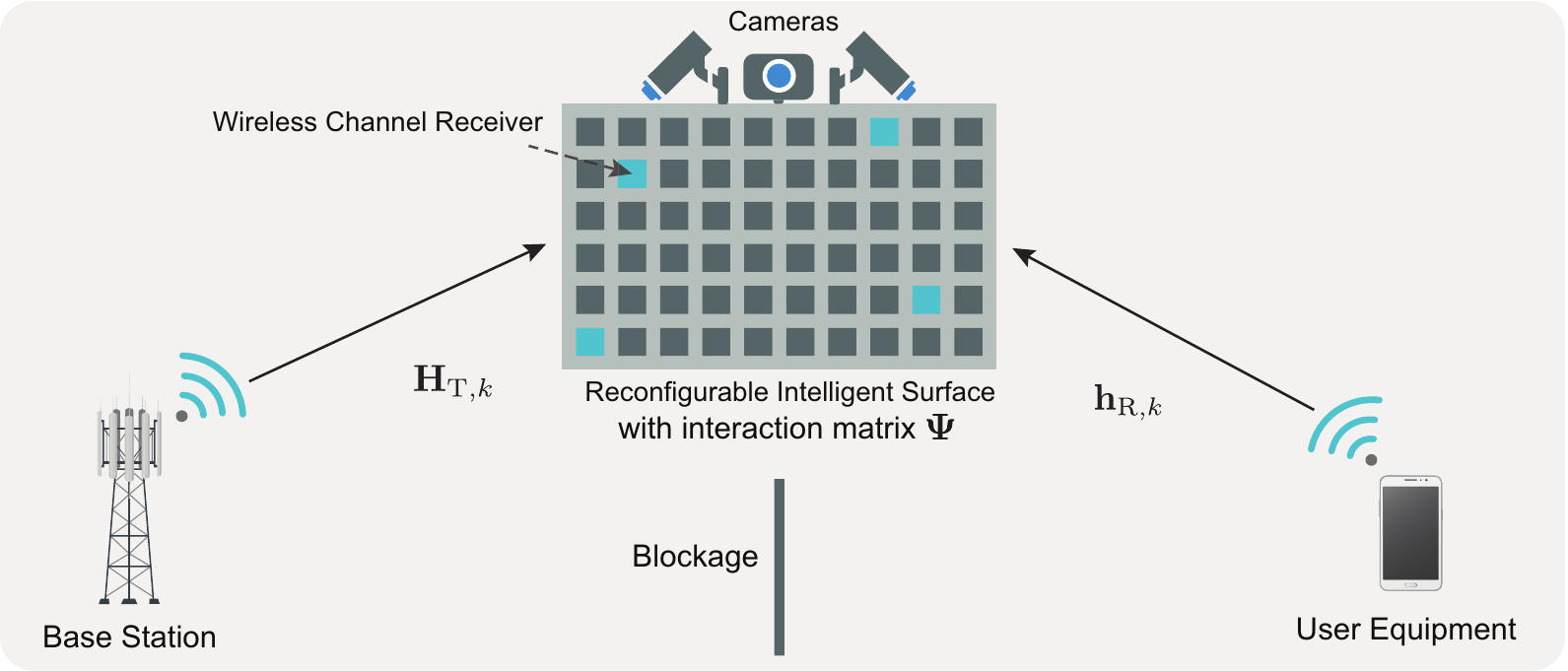}
		\caption{The considered system model consists of a BS, a UE, and an RIS. The link between the BS and the UE is blocked, and their communication is aided by the RIS. The RIS is equipped with cameras and sparsely-distributed receivers, and can intelligently reflect the incident signal to the desired direction by controlling its interaction matrix $\bm \Psi$.}
		\label{fig:system model}
	\end{figure}
	\section{System and Channel Models}\label{System and Channel Models}
	In this section, we describe the system and channel models for the RIS-aided wireless communication scenario considered in this paper.

	\subsection{System Model}
	We consider a wireless communication system where an RIS is placed in the environment to aid the communication between a BS and a UE. For ease of exposition, we assume that a blockage exists between the BS and the UE. Consequently, the BS can only communicate with the UE through the RIS. It is important to mention here, however, that the proposed beam selection in this paper can generally also apply to scenarios with direct links between the BS and the UE. We assume that the BS has an antenna array of $N$ elements, and the UE is equipped with a single antenna. The RIS has $M$ reconfigurable reflecting elements that can shift the phase of the incident signals. Further, the RIS is assumed to be equipped with RGB cameras and sparse wireless channel receivers to obtain sensing information about the surrounding environment. There are three differently-oriented cameras deployed at the center of the RIS surface. These cameras provide a central view and two side views of the surrounding environment. Four reflecting elements at the corners of the RIS are active (connected to baseband), and they act as the wireless receivers~\cite{taha2021enabling}. It is worth noting that the considered system model makes the following assumptions. (i) The RIS lies within the coverage area of only one BS. (ii) The area covered by the RIS cameras is within the coverage area of only one BS, which is the BS that is serving the RIS. (iii) There is only one UE that can be active and served by the RIS.
	\par
	For the uplink and downlink communication, we adopt orthogonal frequency-division multiplexing (OFDM) with $K$ subcarriers. Let $\bH_{{\rm T}, k}\in \bbC^{M \times N}$ and $\bh_{{\rm R}, k}\in \bbC^{M \times 1}$ denote the channel matrix from the BS to the RIS and the channel vector from the UE to the RIS at the $k$-th subcarrier, respectively. If the BS transmits a signal $s_k \in \bbC$ on the $k$-th subcarrier, then, we can write the downlink received signal as
	\begin{align}\label{eq:signal model}
		y_k &= \bh_{{\rm R}, k}^{T} {\bf \Psi} \bH_{{\rm T}, k} \bff s_k + n_k,
	\end{align}
	where $\bff \in \bbC^{N\times 1}$ denotes the beamforming vector of the BS, $\bh_{{\rm T}, k}=\bH_{{\rm T}, k} \bff$ is the effective channel vector between the BS and the RIS (accounting for the BS precoder). The transmitted signal $s_k$ satisfies the power constraint $\bbE\left[s_k^H s_k\right] = \frac{p_t}{K}$ with $p_t$ representing the total transmit power of the BS. $n_k \sim \cN_{\bbC}(0, \sigma^2_n)$ is the complex receive noise at the UE. We use ${\bf \Psi} \in \bbC^{M\times M}$ to denote the RIS interaction matrix which captures the adopted RIS reflection configuration. The matrix ${\bf \Psi}$ is diagonal and can be written as
	\begin{align}
		{\bf \Psi} = \text{diag}({\bm \psi})=\text{diag}(\psi_1, \hdots, \psi_M),
	\end{align}
	where ${\bm \psi}=\left[\psi_1, \hdots, \psi_M\right]^T$ is the diagonal vector of $\bf \Psi$ with $\psi_m \in \bbC$ representing the $m$-th reflecting element of the RIS. $\psi_m$ satisfies $|\psi_m|^2=1$ to capture the constant modulus phase-only constraint. We call $\bm \psi$ the reflecting beamforming vector of the RIS. By configuring the phase shifts of its reflecting elements, the RIS can control the reflected pattern of the incident signal, and help enhance the receive signal power at the UE. The same reflection vector ${\bm \psi}$ is applied to all the K subcarriers due to the time-domain implementation.
	\subsection{Channel Model}
	We adopt a wideband geometric channel model for the channels $\bH_{\mathrm{T},k}$ and $\bh_{\mathrm{R},k}$. With this model, if $\bh_{\mathrm{R},k}$ consists of $L$ clusters, and each cluster $\ell\in[1,L]$ contributes with one ray of time delay $\tau_\ell \in \bbR$, then the delay-d channel vector between the UE and the RIS can be written as
	\begin{equation}\label{eq:delay-d}
		\bh_{\mathrm{R},d} = \sqrt{\frac{M}{\rho}} \sum_{\ell=1}^L \alpha_\ell p(dT_s - \tau_\ell) \ba(\phi^R_{\ell}, \theta^R_{\ell}),
	\end{equation}
	where $\rho$ denotes the pathloss and $p(\tau)$ denotes the pulse shaping function which represents a $T_S$-spaced signaling evaluated at $\tau$ seconds, $\ba(\phi^R_{\ell}, \theta^R_{\ell})$ is the array response vector of the RIS. $\phi^R_{\ell}$ and $\theta^R_{\ell}$ are the corresponding azimuth and elevation angles of arrival (AoA) associated with the $\ell$-th cluster. $\alpha_\ell \in \bbC$ is a complex coefficient of the $\ell$-th cluster.
	\par
	Given the delay-d channel in \eqref{eq:delay-d}, the frequency domain channel vector at subcarrier $k$ can be written as
	\begin{equation} \label{eq-channel}
		\bh_{\mathrm{R},k} = \sum_{d=0}^{D-1} \bh_{\mathrm{R},d} e^{-j\frac{2\pi k}{K} d},
	\end{equation}
	where $D$ represents the maximum delay of the channel corresponding to the delay spread. The channel  $\bH_{\mathrm{T},k}$ is similarly defined.
	\begin{figure}[t]
		\centering
		\includegraphics[width=.7\linewidth]{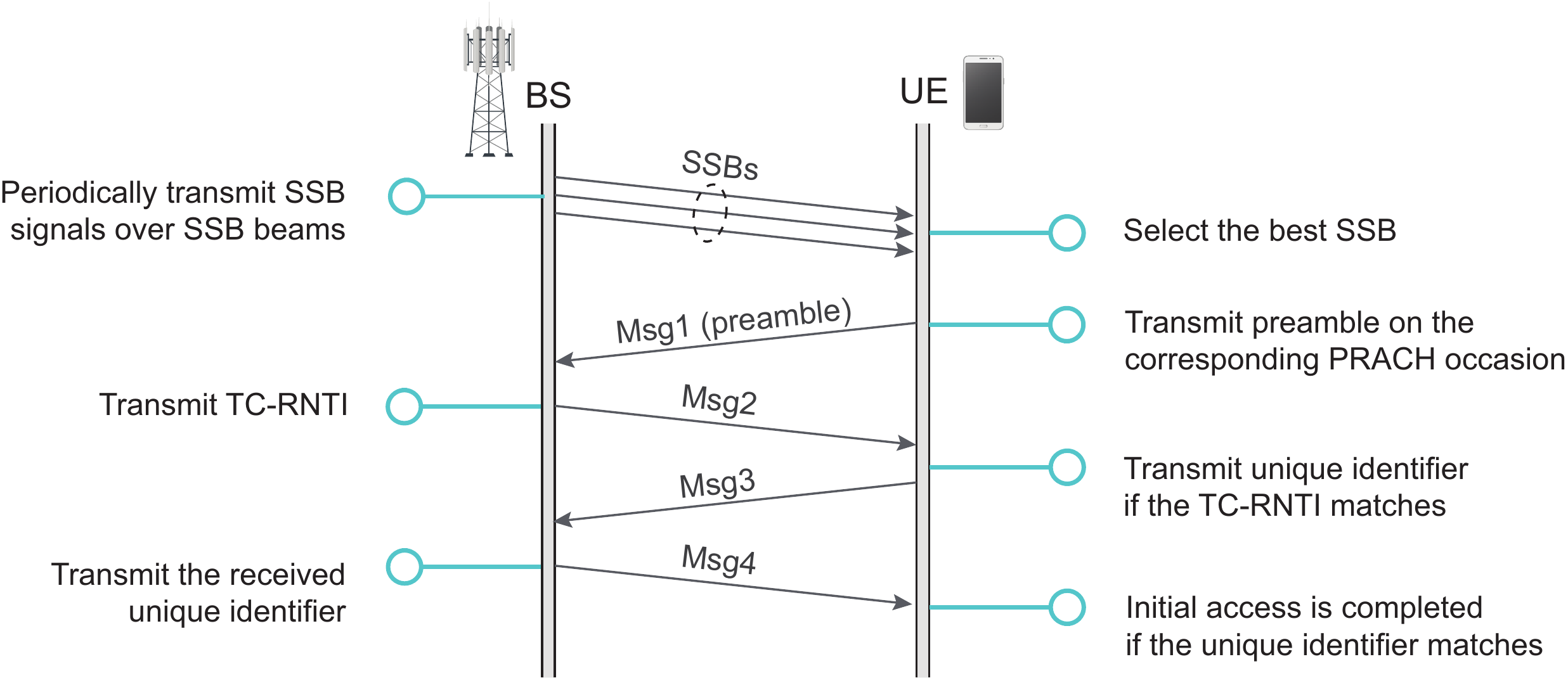}
		\caption{This figure summarizes the signaling and message exchange during the typical 3GPP 5G initial access process.}
		\label{fig:5g_initial_access}
	\end{figure}
	
	\section{3GPP 5G Initial Access: A Brief Background}\label{5G 3GPP Initial Access}
	We assume that the BS and the UE try to initiate a link using the  3GPP 5G protocol \cite{3GPP300}. The objective for the RIS is then to aid the link establishment between the BS and the UE. However, neither the BS nor the UE is aware of the existence of the RIS. To formulate this problem, it is important to first understand how the  3GPP 5G initial access process works. Therefore, in this section, we provide a brief background on the  3GPP 5G initial access process. This process can be summarized in the following steps, which are also illustrated in Fig. \ref{fig:5g_initial_access}.
	\begin{itemize}
		\item \textbf{SSB Signals:} The BS transmits periodic synchronization signal blocks (SSBs) using a predefined set of beams (codebook) \cite{3GPP300}. When a UE wants to access the 5G wireless network, it listens to these SSB signals and blindly decodes them with its set of initial access beams. Based on this beam training process, the UE selects the pair of the SSB beam and receive beam that results in the maximum reference signal receive power (RSRP).
		\item \textbf{Message 1 UE preamble:} After successfully decoding the SSBs, the UE initiates a random access process by transmitting an uplink preamble sequence using the selected receive beam (which maximizes the RSRP of the SSBs) at a physical random access channel (PRACH) occasion. Based on the transmitted preamble sequence and the PRACH occasion where the preamble sequence is transmitted, the BS knows which SSB beam (direction) was selected by the UE \cite{3GPP213}. The preamble sequence is also called Message 1 (Msg1).
		\item \textbf{Message 2 BS random access response:} The BS always listens to the random access channel. Upon detecting a preamble sequence, the BS transmits random access response (RAR) using the same SSB beam. The RAR is also known as Message 2 (Msg2), and it contains a temporary cell radio network temporary identifier (TC-RNTI). The TC-RNTI is calculated based on the PRACH occasion of the preamble sequence. By receiving the TC-RNTI, the UE knows if the preamble sequence it had transmitted was successfully decoded by the BS.
		\item \textbf{Message 3 and 4 Contention resolution:} Finally, Message 3 (Msg3) and Message 4 (Msg4) are exchanged between BS and UE for contention resolution. When multiple UEs detect the same RAR, each of them transmits the Msg3 containing a unique UE identifier. The BS informs only one UE that its random access is completed by transmitting the corresponding unique UE identifier to that UE in Msg4.
	\end{itemize}
	For more details about the initial access process, we refer the reader to \cite{3GPP213}.
	
	\section{Transparent  3GPP 5G RIS: The Key Challenge}\label{5G 3GPP Standalone RIS}
	
	In Section \ref{5G 3GPP Initial Access}, we provided a brief background for the  3GPP 5G initial access process. Now, if the direct link between the BS and UE is blocked, the RIS needs to configure its interaction matrix to aid this communication. But how can this be done if both the BS and the UE do not know about the existence of the RIS? In this section, we provide an initial investigation to address this question and highlight its key challenges.
	\par
	To start, we adopt the achievable rate as the communication performance metric of interest. Given the system model in Section \ref{System and Channel Models} and the downlink receive signal in \eqref{eq:signal model}, the downlink achievable rate for the adopted RIS-based communication system can be written as
	\begin{align}
		R &= \frac{1}{K}\sum_{k=1}^{K}\log_2\left( 1 + \textrm{SNR}\left| \bh^T_{{\rm R}, k} {\bf \Psi} \bH_{{\rm T}, k} \bff \right|^2 \right)\nonumber\\
		&= \frac{1}{K}\sum_{k=1}^{K}\log_2\left( 1 + \textrm{SNR}\left| \left(\bh_{{\rm R}, k}\odot \bH_{{\rm T}, k} \bff \right)^T{\bm \psi}\right|^2 \right),
	\end{align}
	where SNR$=\frac{p_t}{K\sigma_n^2}$ denotes the signal-to-noise ratio.
	The $\odot$ denotes the dot product of two vectors.
	Therefore, for a given BS beamforming vector $\bff$, the optimal reflecting beam for this pair of BS and UE is the one that maximizes the achievable rate as shown by
	\begin{align}\label{eq:best_beam}
		{\bm \psi}^\star = \underset{{{\bm \psi}\in \cO}}{\arg\max} \, \frac{1}{K}\sum_{k=1}^{K}\log_2\left(1 + \textrm{SNR}\left| \left(\bh_{{\rm R}, k}\odot \bH_{{\rm T}, k} \bff\right)^T{\bm \psi}\right|^2 \right),
	\end{align}
	where $\cO$ is the set of all the reflecting beamformers ${\bm \psi}$ that satisfy the constant modulus phase-only constraint, \textit{i.e.}, $|\psi_m|^2=1$. Note that the RIS reflecting beam $\bm \psi$ can be decomposed into the BS-side and the UE-side beam as follows
	\begin{align}\label{eq:decouple_pq}
		{\bm \psi} = {\bp \odot \bq},
	\end{align}
	where $\bp,\bq \in \cO$ denote the BS-side and UE-side beamforming vectors.  With $\bp$ and $\bq$, the optimization problem in \eqref{eq:best_beam} can be equivalently written as
	\begin{equation}\label{eq:best_beam_tmp}
		\left(\bp^\star, \bq^\star\right) = \underset{\substack{\bp \in\cO, \\ \!\!\bq \in\cO}}{\arg\max} \, \frac{1}{K}\sum_{k=1}^{K}\log_2\left(1 + \textrm{SNR}\left| \left(\bh_{{\rm R}, k}\odot \bH_{{\rm T}, k} \bff\right)^T\left(\bp \odot \bq\right)\right|^2 \right).
	\end{equation}
	
	\begin{figure}[t]
		\centering
		\subfigure[BS-side beam selection]{\includegraphics[width = 0.325\linewidth]{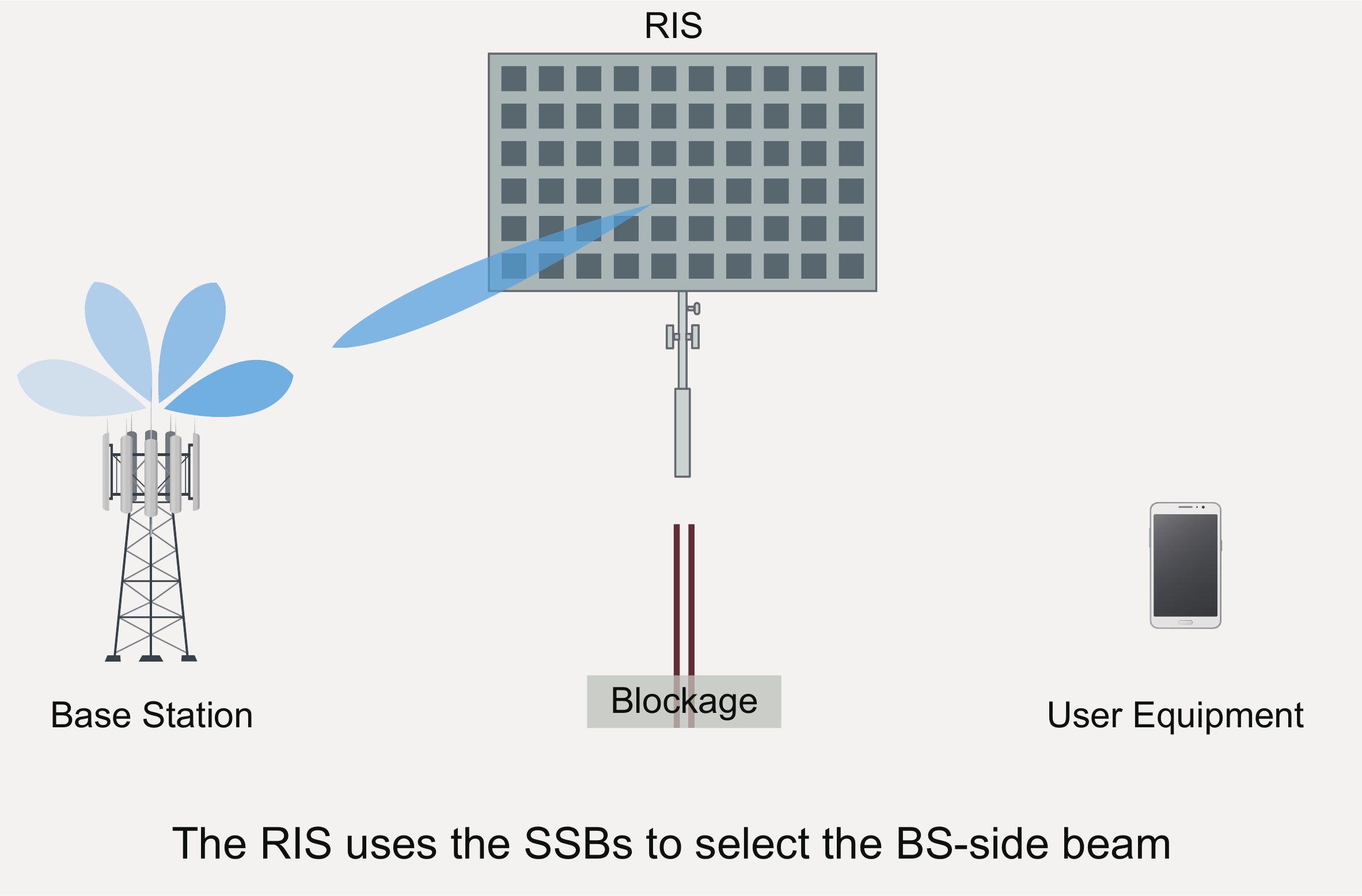}}
		\subfigure[Key challenge]{\includegraphics[width = 0.325\linewidth]{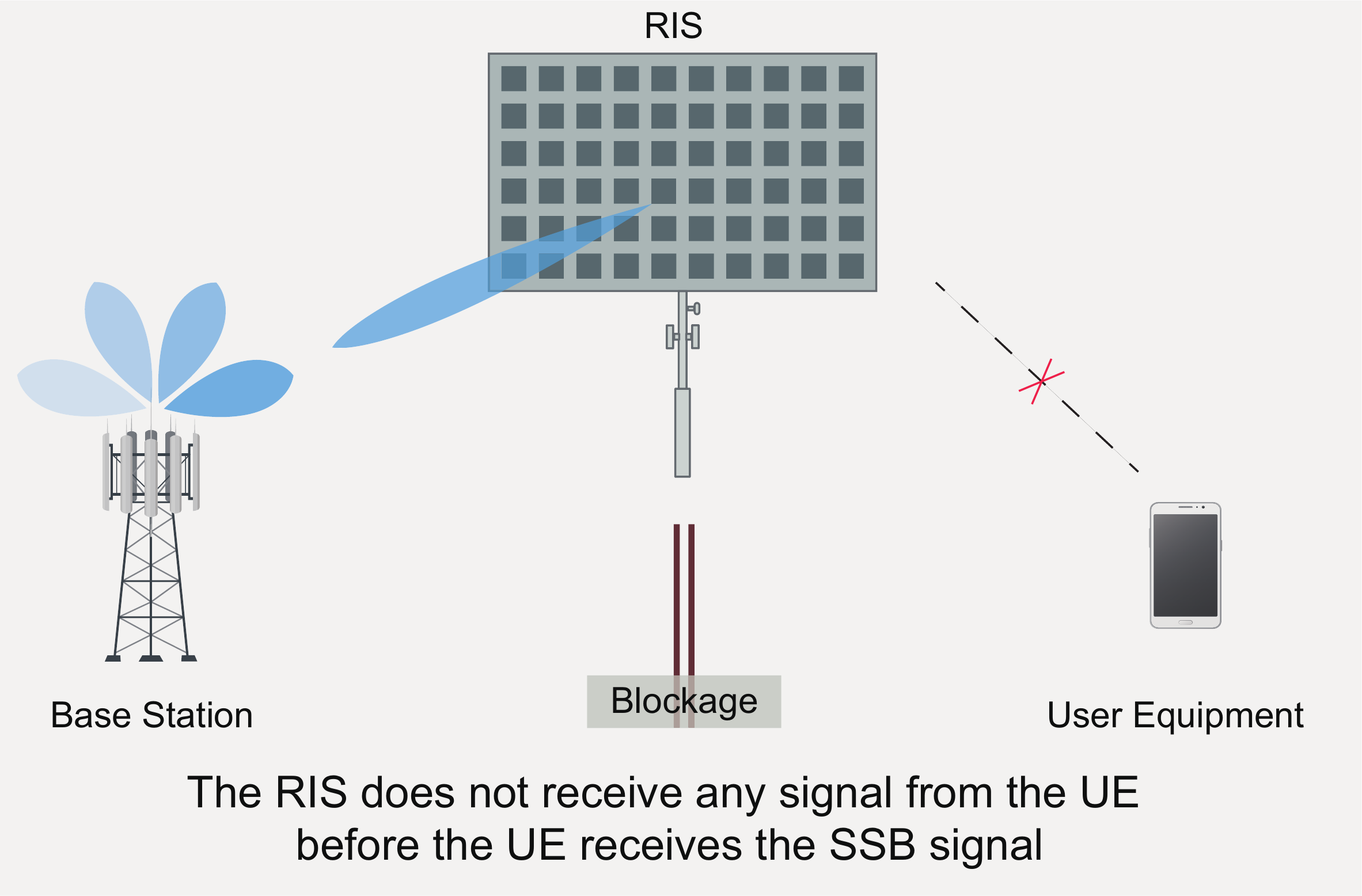}}
		\subfigure[UE-side beam sweeping]{\includegraphics[width = 0.325\linewidth]{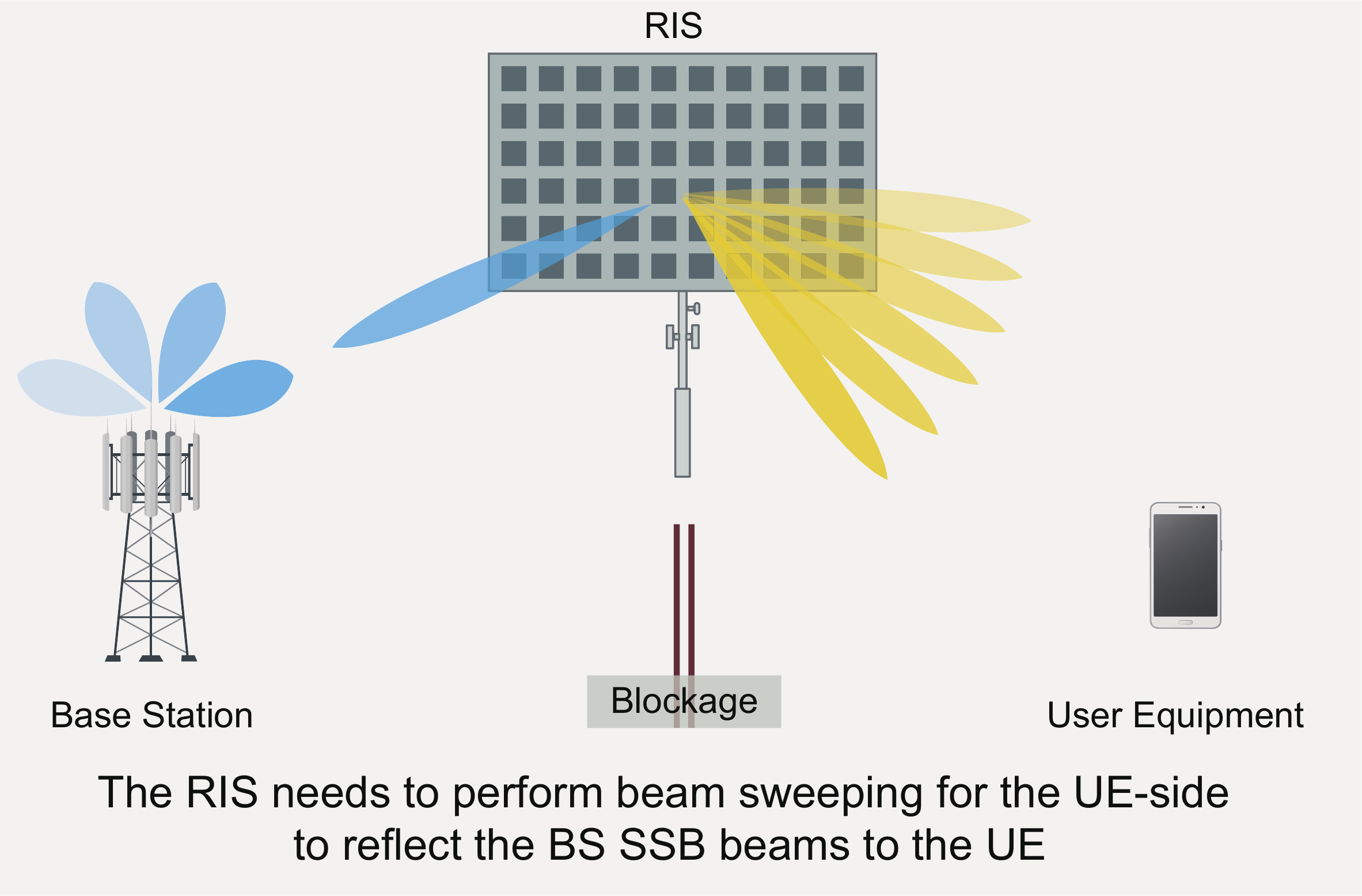}}
		
		\caption{This figure illustrates the key challenge in applying the 3GPP 5G initial access process for an RIS-aided communication system; the UE only sends a preamble sequence after it receives and decodes the SSBs. Hence, the RIS cannot have any information about the UE channel and it needs to perform beam sweeping over a very large codebook.}
		\label{fig:beam_training}
	\end{figure}

	Now, to account for the practical constraint of quantized phase shifters \cite{Heath2016}, we limit the search space of $\bp$ and $\bq$ to pre-designed finite-size codebooks $\cP$ and $\cQ$. It is important to note here that (the new optimization problem) can act as an upper bound of the designed codebooks.
	Given \eqref{eq:decouple_pq} and the two codebooks, the optimization problem in \eqref{eq:best_beam} can be re-written as
	\begin{equation}\label{eq:best_beam_decouple}
		\left(\bp^\star, \bq^\star\right) = \underset{\substack{\bp \in\cP, \\ \!\!\bq \in\cQ}}{\arg\max} \, \frac{1}{K}\sum_{k=1}^{K}\log_2\left(1 + \textrm{SNR}\left| \left(\bh_{{\rm T}, k}\odot \bp\right)^T \left(\bh_{{\rm R}, k} \odot \bq\right) \right|^2 \right).
	\end{equation}
	It can be seen from \eqref{eq:best_beam_decouple} that the optimal BS-side beam $\bp^\star$ and the optimal UE-side beam $\bq^\star$ depend on both the BS-side channel $\bh_{{\rm T}, k}$ and the UE-side channel $\bh_{{\rm R}, k}$. To further simplify the RIS operation, we decouple the optimization problem in \eqref{eq:best_beam_decouple} to the sub-optimal approximation as follows:
	\begin{subequations}
		\small
		\label{eq:best_beam_easy_easy}
		\begin{empheq}[left={\empheqlbrace}]{align}
			& {\bp}^\star = \underset{\bp \in \cP }{\arg\max}\frac{1}{K}\sum_{k=1}^{K}\left| \left(\bh_{{\rm T}, k}\odot \bp\right)^H \ba^* \right|^2 \label{eq:best_beam_easy_a}\\
			&{\bq}^\star = \underset{\bq \in \cQ }{\arg\max}\frac{1}{K}\sum_{k=1}^{K}\left| \left(\bh_{{\rm R}, k}\odot \bq\right)^H \ba \right|^2 , \label{eq:best_beam_easy_b}
		\end{empheq}
	\end{subequations}
	where $\ba \in \bbC^{M\times 1}$ is an arbitrary reference vector. It is worth noting that the $\bp^\star$ and $\bq^\star$ obtained in \eqref{eq:best_beam_easy_easy} is the \textit{optimal} solution to \eqref{eq:best_beam_decouple} for a very important case, \textit{i.e.}, when the BS-side channel and the UE-side channel only contain the line-of-sight (LoS) path, and $\cP=\cQ=\cO$. We provide the proof in Appendix~\ref{Appendix}.
	\par
	\textbf{BS-side RIS beam selection:} To obtain the optimal BS-side beamforming vector $\bp^\star$ in \eqref{eq:best_beam_easy_a}, some information on the BS is needed. To that end, we propose that the RIS can observe the BS by exploiting the SSBs as pilot signals. First, the RIS blindly decodes the SSBs with its sparse wireless channel receivers, and synchronizes with the BS. Second, the RIS uses the previously decoded SSBs as predefined pilot signals, and estimates the channels between the BS and the sparse wireless channel receivers. The channel estimates at the sparse wireless channel receivers draw a defining multi-path signature not only of the BS location, but also of its surrounding environment \cite{alkhateeb2018deep}. Finally, the RIS uses an offline-trained DL model to predict the BS-side beamforming vector $\bp^\star$ from this signature. Note that the RIS does not need any extra/dedicated signaling from the BS to design $\bp^\star$ because the SSBs are always periodically transmitted by the BS (with or without the RIS).
	\par
	\textbf{The Key Challenge:} Selecting the UE-side beamforming vector from the codebook $\cQ$ is the key challenge. According to the 5G 3GGPP initial access introduced in Section \ref{5G 3GPP Initial Access}, the UE does not transmit any signal/preamble until it detects the BS SSBs signals. Furthermore, when there is a blockage between the BS and the UE, the UE may not be able to detect the BS SSBs without a communication link established by the RIS. \textbf{Therefore, the RIS is required to configure its interaction matrix to guarantee sufficient receive signal power at the UE \textit{before} the UE can detect any signal from the BS, and before it (the RIS) receives any signals from the UE.} In other words, the RIS needs to configure the interaction matrix \textit{without} receiving any signals from the UE. The trivial solution to design the UE-side beamforming vector at the RIS is the exhaustive search (beam sweeping) over all the beams in the codebook $\cQ$ until the UE responds with an index corresponding to one of the beams. However, the number of beams that the RIS supports generally grows proportionally to the number of reflecting elements to fully exploit them. \textbf{Due to the large number of reflecting elements at the RIS, the beam sweeping requires tremendous training overhead.} Further, the beam sweeping becomes more impractical when the UE is mobile as the beam sweeping needs to be frequently repeated in short periods of time.
	\par
	Then, our objective is to solve the optimization problem in \eqref{eq:best_beam_easy_b}, \textit{i.e.}, finding the UE-side beamforming vector $\bq^\star \in \cQ$, with the smallest number of trials. In Section \ref{Sensing for Standalone RIS Operation}, we propose to leverage sensing to achieve this objective.

	\section{Sensing for Standalone RIS Operation}\label{Sensing for Standalone RIS Operation}
	The convergence of communication, sensing, and localization is considered one of the key features in 6G and beyond \cite{rappaport2019wireless,de2021convergent}. The sensing and localization capabilities may not just support new interesting applications such as AR/VR and autonomous driving, but also provide rich information and awareness about the surrounding environment to the communication systems. This could be particularly useful for mmWave/THz systems where the communication links are mainly line-of-sight (LoS) and highly dependent on the geometry of the surrounding environment and the locations of the transmitters/receivers. In this section, we propose to utilize sensing at the RIS to build up environment-awareness and leverage this awareness in enabling efficient transparent operation for these surfaces. Next, we briefly explain the key idea of the proposed sensing-aided RIS operation in \sref{Key Idea: Observe with Sensing} and \sref{RIS Beam Set Prediction}. Then, we describe the proposed vision-aided transparent RIS  3GPP 5G operation in Section \ref{Vision-Aided Transparent RIS for 5G 3GPP}.
	\subsection{Key Idea: Observe with Sensing}\label{Key Idea: Observe with Sensing}
	The key idea of the proposed approach can be summarized as follows: If the RIS is equipped with some sensors (such as radar, LiDAR, camera, etc.), these sensors could provide useful information about the surrounding environment. This information, such as the geometry of the scatterers and the locations/directions of the transmitters/receivers can help the RIS-integrated wireless communication system in several ways. (i) The RIS can leverage the sensing information for identifying the promising beamforming directions and avoid extensive (blind) beam training. (ii) The sensing information can potentially help the RIS manage its beams to avoid causing interference to adjacent users associated with neighboring BSs. (iii) By periodically monitoring the locations/directions of the candidate UEs, the RIS can track the UEs' beams and model their mobility patterns. Next, we focus on the first potential gain, which is reducing the beam training overhead.
	
	\subsection{RIS Beam Set Prediction}\label{RIS Beam Set Prediction}
\begin{figure*}[t]
		\centering
		\includegraphics[width=1\linewidth]{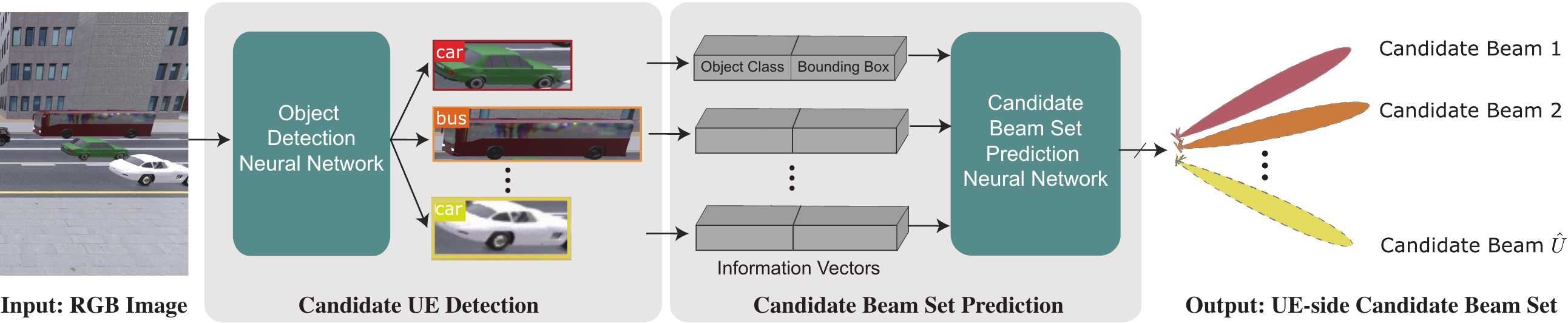}
		\caption{Overview of the proposed ML framework for the RIS UE-side candidate beam set prediction. Given the visual information of the surrounding environment of the RIS, the framework first detects the candidate UEs with an object detector. Then the information about the candidate UEs (bounding boxes and object classes) are used to predict the candidate beams.}
		\label{fig:framework}
	\end{figure*}

	As discussed in Section \ref{5G 3GPP Standalone RIS}, the key challenge of realizing standalone RIS operation lies in the high beam training overhead associated with UE-side RIS beam sweeping. To tackle this problem, we propose an ML framework that (i)~utilizes visual sensors at the RIS to obtain information about the candidate UEs in the scene, and (ii)~exploits this information to reduce the UE-side beam training overhead of the RIS. An overview of the proposed ML framework is highlighted in Fig. \ref{fig:framework}. With the equipped cameras, the RIS first obtains visual information, namely, RGB images, of the surrounding environment. Then, the RIS leverages computer vision object detection models to identify the candidate UEs in the scene. The information of all the detected candidate UEs is then used to predict the set of the most promising beams for the candidate UEs. Note that the number of these promising/candidate beams is proportional to the number of candidate UEs in the scene, which is typically much smaller than the size of the RIS beam codebook. This highlights the potential of significantly reducing the beam training overhead by leveraging the available visual information.
	\par
	The objective of the ML framework is then to accurately predict a set of UE-side candidate beams that correspond to the candidate UEs in a given scene. More specifically, for each UE, the objective is to accurately predict the UE-side RIS beam $\bq^\star$ that satisfies \eqref{eq:best_beam_easy_b}. The optimal UE-side candidate beam set of an image can then be defined as
	\begin{align}\label{eq:best_beamset}
		{\bQ}^\star =\left\{{\bq}_1^\star, \hdots, {\bq}_U^\star \right\},
	\end{align}
	where $U$ is the total number of ground-truth candidate UEs in the image, and ${\bq}_u^\star\ (u=1,\hdots, U)$ is the optimal UE-side beamforming vector for the $u$-th ground-truth UE. The optimal framework $f^\star(\cdot)$ is then defined as the one which can perfectly predict the optimal UE-side candidate beam set for any given image. Let $\bX \in \bbR^{w\times h \times 3}$ denote the input RGB image to the ML framework with $w$ and $h$ denoting the width and height of $\bX$. The optimal framework can then be expressed as
	\begin{align}\label{eq:framework}
		f^\star(\bX) ={\bQ}^\star_\bX,
	\end{align}
	where ${\bQ}^\star_\bX$ is the optimal UE-side candidate beam set of image $\bX$. Deriving the exact expression of $f^\star(\cdot)$ is very difficult since it depends on the channel model, the visual model, and the layout of the environment around the UE and RIS.
	This motivated leveraging deep learning models to learn the complex function $f^\star(\cdot)$ in a data-driven manner. The adopted DL models will be explained in detail in Section~\ref{Deep Learning Modeling}.
	\subsection{Vision-Aided Transparent RIS for  3GPP 5G}\label{Vision-Aided Transparent RIS for 5G 3GPP}
\begin{figure*}[t]
		\centering
		\includegraphics[width=1\linewidth]{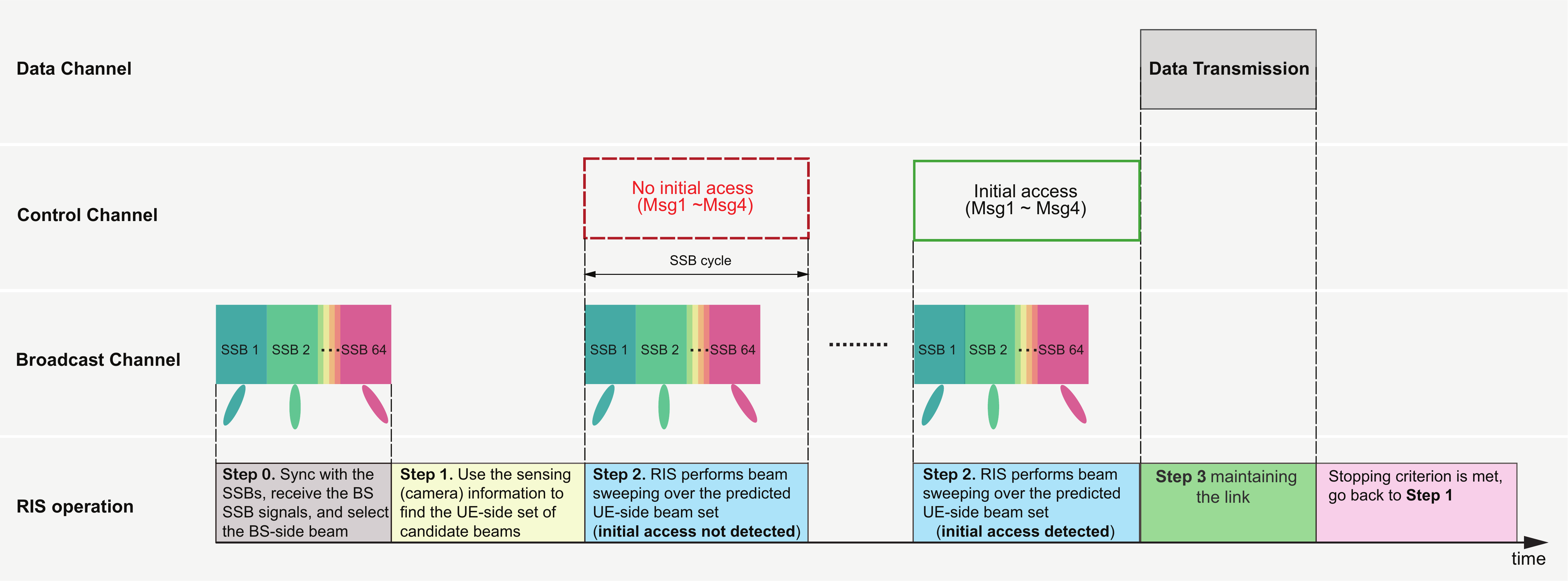}
		\caption{This figure demonstrates the proposed vision aided standalone RIS system operation. Exploiting the visual information, the ML framework reduces the UE-side beam training overhead for the RIS (step 1). The timing of each step of the proposed RIS beam selection shows its compliance with the 3GPP 5G protocol. }
		\label{fig:vision-aided beam prediction}
	\end{figure*}

	Given the general idea and motivation of the proposed sensing-aided RIS operation presented in \sref{Key Idea: Observe with Sensing}, we next describe in detail the proposed transparent  3GPP 5G  operation of the vision-aided RIS system. Consider the 3GPP initial access procedure described in \sref{5G 3GPP Initial Access}, the following steps (illustrated in Fig. \ref{fig:vision-aided beam prediction}) summarize how the vision-aided RIS can efficiently operate as part of this initial access process:
	\begin{itemize}
		\item \textbf{Step 0. RIS predicts the BS-side beam:} Using the sparse wireless channel receivers, the RIS blindly decodes the SSBs which are periodically transmitted by the BS. These SSBs are first used to synchronize with the BS. Then the RIS exploits the SSBs as pilot signals to draw the multi-path signature of the BS. From this BS multi-path signature, the RIS obtains the BS-side beamforming vector $\bp^\star$ in \eqref{eq:best_beam_easy_a} with a offline-trained DL model. Note that, we assume the RIS knows beforehand the initial access configuration including the SSB periodicity, bandwidth, and carrier frequency. This can be achieved if the RIS has the same capability as a UE device. Note that, we assume that the RIS lies within the coverage area of only one BS, otherwise it may not be able to correctly align its BS-side beam with the intended BS.
		\item \textbf{Step 1. RIS predicts UE-side candidate beam set with camera information}: The RIS obtains the visual information (images) of the surrounding environment with its visual sensors. Given this visual information, the RIS detects the candidate number of UEs $\hat{U}$ and predicts the corresponding UE-side candidate beam set $\hat{\bQ}$ with the ML framework shown in Fig. \ref{fig:framework}. Note that, we assume that the area covered by the images is within the coverage area of only one BS, which is the BS that the RIS is aligned with in Step 0.
		\item \textbf{Step 2. RIS performs beam sweeping over the predicted UE-side beam set}: The RIS applies the BS-side beam vector obtained in Step~0, and performs beam training within the UE-side candidate beam set $\hat{\bQ}$ obtained in Step 1. For each UE-side candidate beam $\hat{q}_u \in \hat{\bQ}$, the RIS holds the beam for a time window longer than one SSB cycle (\textit{e.g.} 20 milliseconds). Meanwhile, using its sparse wireless channel receivers, the RIS tries to detect a successful initial access (Msg1 $\sim$ Msg4) by detecting the signal power on the wireless bands where the initial access messages are transmitted\footnote{By monitoring the signal power on the wireless band, the RIS does not need to decode the messages exchanged between the BS and the UE in the initial access process.}. If the RIS cannot detect a successful initial access, it concludes that no UE can use the currently tested UE-side beam. The RIS then switches to the next UE-side candidate beam in the set $\hat{\bQ}$, and repeats the same process. As shown by Fig. \ref{fig:vision-aided beam prediction}, when testing the UE-side beams, the RIS ensures its beam switching is synchronized with the BS's SSB beam sweeping. In such a manner, the UE can have a chance to receive one or more complete SSBs. Note that the proposed transparent RIS operation focuses on the single-user MIMO scenario. The RIS may detect multiple candidate UEs in \mbox{Step 1}, however, it serves only one of the candidate UEs. We briefly discuss the multi-user scenario in \sref{Future Work}.
		\item \textbf{Step 3. Maintaining the link}: 
		If the RIS detects a successful initial access, it indicates that one UE is communicating with the BS via the current RIS beam. The RIS should perform beam tracking (starting from this beam) to consistently serve this UE as it moves. To achieve beam tracking, the sensing information obtained at the RIS can be exploited to infer the optimal UE-side beam for the current time instance. Furthermore, enabled by the sensing capability, the RIS can predict the dynamics of the environment and the UE to achieve proactive beam tracking (predict the future optimal UE-side beam). During the beam tracking, the RIS keeps monitoring certain stopping criteria. After a stopping criterion is met, the RIS stops beam tracking and goes back to Step 1.
		\item \textbf{Stopping criteria:}
		We expect that the RIS beam tracking stops when the communication between the UE and the BS is terminated. More specifically, the RIS beam tracking stopping criteria include the following: (i) There is a blockage between the RIS and the BS/UE. (ii) The BS has switched to serve another UE. (iii) The communication session between the BS and the UE is terminated. We discuss more details about the RIS beam tracking and the stopping criteria in \sref{Future Work}.
	\end{itemize}
	\par
	Fig. \ref{fig:vision-aided beam prediction} shows the timing of each step of the proposed vision aided standalone RIS system operation with respect to the 3GPP 5G initial accessed process. This demonstrates the compatibility of the proposed beam selection with the 3GPP 5G protocol.
	\section{Deep Learning Modeling}\label{Deep Learning Modeling}
	The proposed sensing-aided transparent RIS approach, presented in \sref{Sensing for Standalone RIS Operation}, depends on the capability of the RIS to leverage its sensors (cameras in this paper) to determine the set of candidate beams. In this paper, we propose to employ the powerful learning capabilities of computer vision and deep learning to achieve this task, which we divide into two sub-tasks, namely candidate UE detection in the field of view, and candidate beam set prediction. In this section, we describe the adopted ML models for these two sub-tasks.
	\subsection{Candidate UE detection}\label{Candidate UE detection}
	Since convolutional neural networks (CNNs) surpassed human-level performance on an image classification task for the first time\cite{he2015delving}, it has attracted increasing attention and is underway to being utilized in a variety of computer vision tasks. Object detection, as a classical computer vision task, has been extensively studied with CNN-based DL models \cite{yolov3, ren2016faster,tan2019efficientnet}. Therefore, we adopt a CNN-based object detection model in this paper. Since the wireless environment is typically changing quickly, the object detector in the proposed framework for UE-side candidate beam set prediction needs to satisfy an essential requirement: the capability to produce fast object detections of high quality. To that end, we select the well-known YOLOv3 object detector \cite{yolov3} because of its fast prediction speed and high accuracy. The YOLOv3 also has advantages including easy implementation and good compatibility with the hardware in the industry, which makes it more applicable in real-world deployments. To obtain a YOLOv3 model that can accurately detect candidate UEs with the visual information at the RIS, there is no need to train the model from scratch. Instead, we start from a pre-trained model and fine-tune it on the images from the RIS's cameras. Given an input image, the YOLOv3 model outputs a class index $c\in \bbZ$, and a bounding box vector $\bb\in \bbR^{4 \times 1}$ for each detected object. The bounding box vector consists of the x-center, the y-center, the width, and the height of the bounding box. We refer readers to \cite{yolov3} for more details of the YOLOv3 model.
	\subsection{Candidate Beam Set Prediction} \label{Candidate Beam Set Prediction}
	In Section \ref{Candidate UE detection}, we discuss the process of using the YOLOv3 object detector to obtain classes and bounding boxes of the candidate UEs in an image. Based on this information, we now design a NN that can predict the UE-side candidate beam set. Next, we will describe the key components of the proposed NN for the UE-side candidate beam set prediction, namely the input/output representation, the NN architecture, and the loss function and learning model.
	\par
	\noindent \textbf{Input/Output Representation and Normalization:}
	Given one image, the YOLOv3 model detects $\hat{U}$ candidate UEs. For each candidate UE, the YOLOv3 model outputs a class index $c\in \bbZ$, and a bounding box $\bb\in \bbR^{4 \times 1}$. To make the training process of the NN faster and more stable,
	we convert the class $c$ to a one-hot vector $\bar{\bc}$, and we normalize the bounding box $\bb$ by the size of the image, $w$ and $h$. The normalized bounding box is denoted by $\bar{\bb}$. Then the one-hot representation of the class $\bar{\bc}$ and the normalized bounding box $\bar{\bb}$ are concatenated to the candidate UE information vector $\bv = [{\bar{\bc}}^T, \bar{\bb}^T]^T$. Finally, the input matrix $\bV$ to the NN architecture is written as
	\begin{align}
		\bV = [\bv_1, \hdots, \bv_{\hat{U}}, {\bf 0}, \hdots, {\bf 0}].
	\end{align}
	Note that we pad $(U_{max}-\hat{U})$ zero-vectors since the number of detected UEs varies from image to image. $U_{max}$ denotes the maximum number of candidate UEs that exist in any image.
	\par
	To construct the desired output of the NN, we first obtain the optimal UE-side beam set $\bQ^\star$ corresponds to the image as shown by \eqref{eq:best_beamset} with exhaustive search over the codebook $\cQ$. Then we convert $\bQ^\star$ into a multi-hot vector $\bt^\star=\left[t_1^\star,\hdots, t^\star_{|\cQ|}\right]$. For any element of $\bt^\star$, $t_j^\star$ satisfies
	\begin{align}\label{thresh}
		t_j^\star=
		\begin{cases}
			1 & \cQ_{j} \in \bQ^\star \\
			0 & \text{otherwise},
		\end{cases}
	\end{align}
	where $\cQ_{j}$ is the $j$-th beam in $\cQ$.
	
	\par
\begin{figure}[t]
		\centering
		\includegraphics[width=0.55\linewidth]{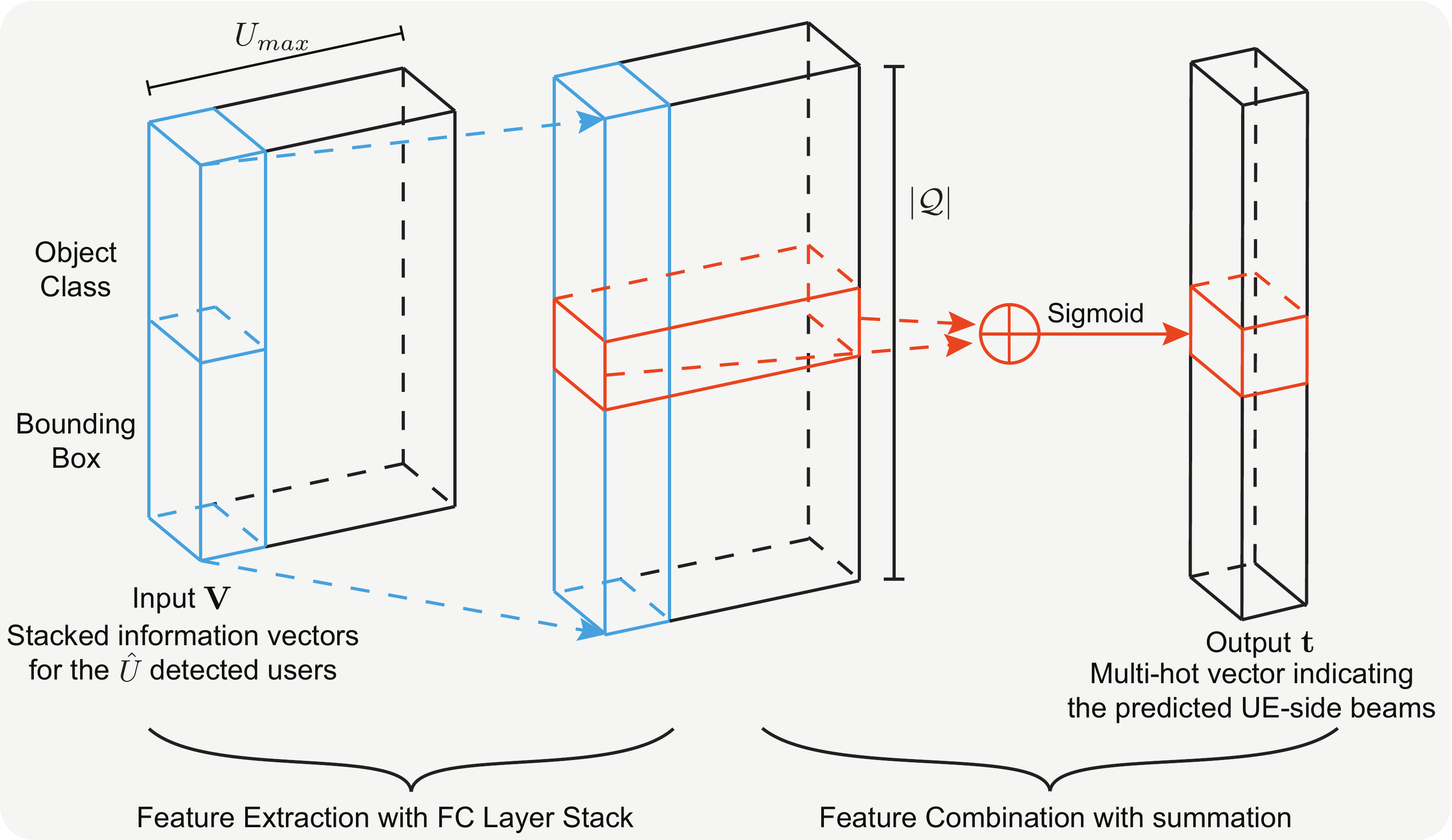}
		\caption{The proposed NN architecture for predicting candidate beam set from detected UE information. The NN architecture first applies the same stack of FC layers on all the UE information vectors to project them into $|\cQ|$-dimensional vectors. The $|\cQ|$-dimensional vectors are then added up and activated by the sigmoid function.}
		\label{fig:NN_arch}
	\end{figure}

	\noindent \textbf{NN Architecture:}
	As shown by Fig. \ref{fig:NN_arch}, we propose an NN architecture that effectively predicts the UE-side candidate beams from the class and bounding box information of all detected candidate UEs in an image. The proposed NN architecture first applies the same stack of fully connected NN layers on each candidate UE's information vector (each column of $\bV$) to extract high-level features. These fully-connected layers adopt the ReLU activation function given by $f_{\text{ReLU}}(x)=\max(x, 0)$. After this feature extraction, each input candidate UE information vector is transformed to a $|\cQ|$-dimensional vector. Then these $|\cQ|$-dimensional vectors are combined by a summation operation. Since the desired output of the NN is designed as a multi-hot vector, the sigmoid activation is applied to the combined vector to restrict its value into the range $(0,1)$. After the sigmoid activation, we obtain the output vector, $\bt \in \bbR^{|\cQ|\times 1}$.
	\par
	The proposed NN architecture has two advantages for predicting the UE-side candidate beam set from the candidate UE information vectors.
	\begin{itemize}
		\item First, it reuses the same stack of fully connected layers to extract features from different candidate UE information vectors. This aligns with the intuition that all candidate UEs are equivalent for the NN architecture, thus, they should be processed in the exact same way. Reusing this same stack of fully connected layers also reduces the complexity of the proposed NN architecture, which stabilizes the training process and reduces the computational complexity for the inference process.
		\item Second, the output of the proposed NN architecture does not rely on the order of the input candidate UE information vectors. This is achieved by reusing the same stack of fully connected layers and the summation operation that combines features from all candidate UE information vectors. Therefore, the proposed NN architecture can be more robust by not overfitting to the order of the input candidate UE information vectors.
	\end{itemize}
	We validate our intuitions on the NN structure by numerical results presented in \sref{sim result1}.
	\par
	\noindent \textbf{Loss Function and Learning Model:}
	The NN is designed to predict the UE-side candidate beam set based on the candidate UEs detected by the YOLOv3 model. This can be modeled as a multi-class classification problem. Therefore, we adopt a classification learning model. We train the NN by supervised learning and employ the cross-entropy loss function expressed by
	\begin{align}
		L_{CE}(\omega_{NN}) = - \sum_{i=1}^{N_{tr}} \sum_{j=1}^{|\cQ|}t^{\star}_{i,j}\log_2(t_{i,j}),
	\end{align}
	where $t_{i,j}^{\star}$ is the $j$-th element of the \textit{desired} output vector $\bt_{i}^{\star}$ of the $i$-th training sample, and $t_{i,j}$ is the $j$-th element of the NN's output vector of the $i$-th training sample. $N_{tr}$ is the number of training samples. $\omega_{NN}$ denotes the trainable parameters of the NN.
	\section{Dataset and Performance Metrics}\label{setup}
	In this section, we first explain in detail the considered simulation setup. Second, we elaborate on the generation process of the dataset, which is later used to train and evaluate the NN. Then, we introduce the metrics used to evaluate the UE-side beam set prediction performance.
	\subsection{Simulation Setup}
\begin{figure}[t]
		\centering
		\includegraphics[width=.65\linewidth]{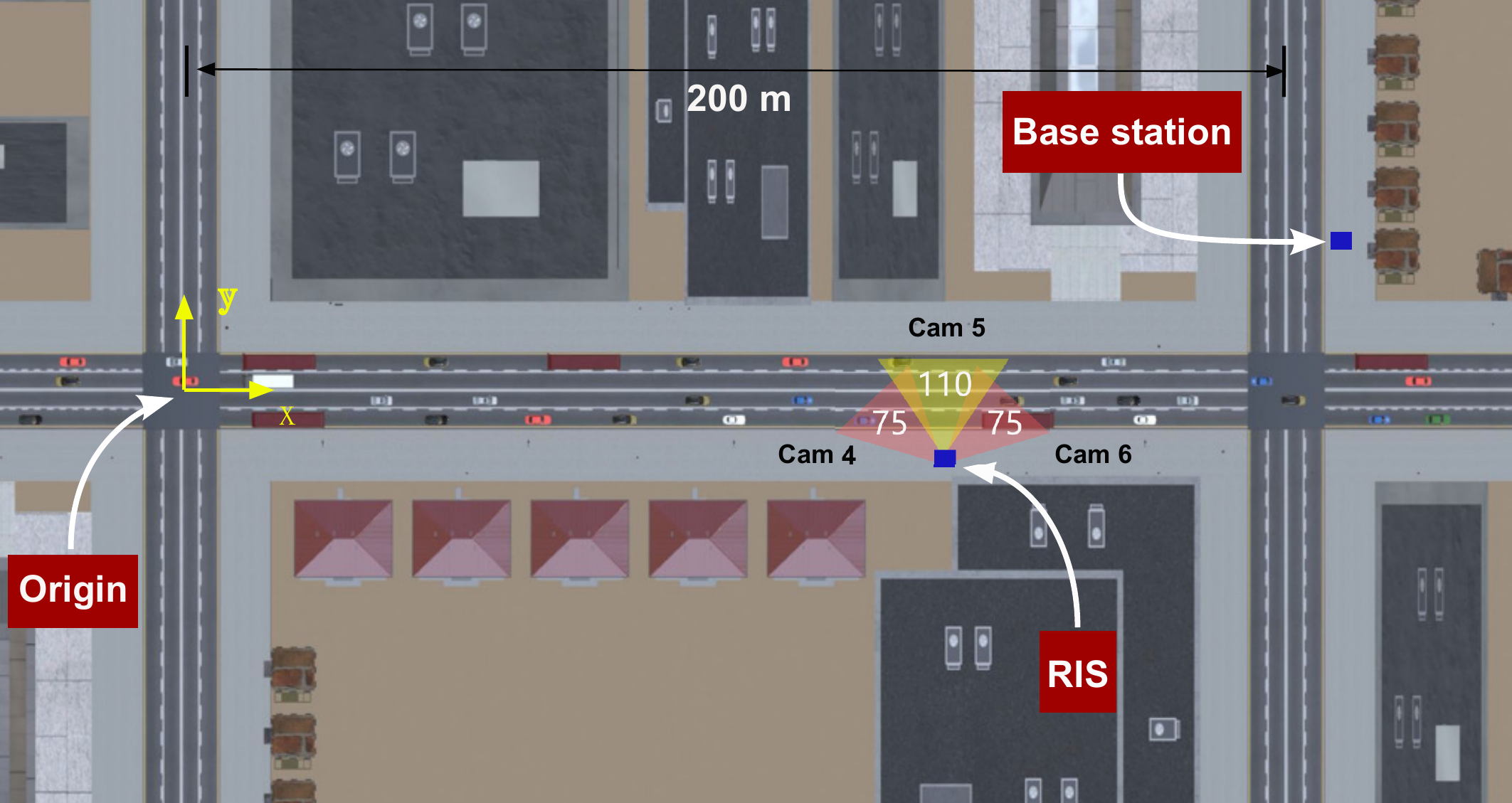}
		\caption{A top-view of the adopted simulation scenario of the ViWi dataset. This scenario models a busy downtown area with a variety of moving and stationary objects including cars, buses, trucks, trees, and buildings.}
		\label{fig:viwi}
	\end{figure}

	We investigate utilizing sensing based perception to enable beamforming for standalone RIS. Hence, realistic wireless and visual modelings are essential for our simulation. To that end, we generate the training and test data with the ViWi dataset \cite{viwi}. The ViWi dataset provides co-existing wireless and visual data based on accurate ray-tracing. It comprises sequences of RGB frames, beam indices, and user link statuses. They are generated from a large simulation of a synthetic outdoor environment depicting a downtown street with multiple moving objects. To simulate our RIS-aided communication system, we construct a new scenario based on the ViWi scenario 1. A top view of our scenario is presented in Fig. \ref{fig:viwi}. The BS is located on the vertical street at the upper right. The UEs are the moving vehicles on the main street. The RIS is installed at the side of the main road to aid communication between the BS and the UEs. In the simulation, for simplicity, we assume the BS and the UEs to be single-antenna. The RIS is equipped with a uniform planar array (UPA) with 32 columns and 8 rows of reflecting elements, \textit{i.e.}, the total number of reflecting elements of the RIS is $256$. Three cameras (``Camera 4", ``Camera 5" and ``Camera 6") are deployed at the RIS as shown in Fig. \ref{fig:viwi}. The central ``Camera 5" has a $110^\circ$ field of view while the side cameras' field of views are $75^\circ$. In the following simulations, we focus on the UEs in the views of ``Camera 4" and ``Camera~5".
	\subsection{Data Generation}
	We first generate 10000 scenes with the ViWi dataset. Then, for ``Camera 4", each scene consists of an image $\bX$, and the channels of all UEs in the view of the camera. Note that we only keep the data where the UEs do not have LoS paths with the BS. We obtain the optimal UE-side candidate beam set ${\bQ}^\star$ for each image according to \eqref{eq:best_beam_easy_b} and \eqref{eq:best_beamset} by exhaustively searching over the codebook $\cQ$. The optimal UE-side candidate beam set for each image is then converted to the multi-hot representation $\bt^\star$. Then $\bX$ and $\bt^\star$ form a data point $(\bX, \bt^\star)$. We apply the same process to the data from ``Camera 5". After processing all 10000 scenes, we have a dataset for ``Camera 4" with 9384 data points, and a dataset for ``Camera 5" with 5955 data points.
	\par
	As discussed in \ref{Candidate UE detection}, to fine-tune the YOLOv3 model, 500 images are randomly selected as the training data for each camera. We manually label the chosen images with bounding boxes and object classes for all visible vehicles. Then we fine-tune the YOLOv3 model on these manually labeled images.
	\par
	After fine-tuning the YOLOv3 model, we generate the datasets to train the beam set prediction NNs. First, the fine-tuned YOLOv3 model is applied to all images in the two datasets to obtain the candidate UE information vectors. After that, each data point $\left(\bX, \bt^\star\right)$ in the two datasets is converted to $\left(\bV, \bt^\star\right)$. The two datasets for ``Camera 4" and ``Camera 5" are split into training and test sets using an $80\%$-$20\%$ data split. Note that two NNs are separately trained and evaluated on the two datasets for ``Camera 4" and ``Camera~5" since they have different camera angles.
	\subsection{Performance Mectrics} \label{subsec:metrics}
	Here, we present the metrics followed to evaluate the quality of the NNs' UE-side beam set prediction. In the NNs' output vector $\bt$, each element represents a promising score of the corresponding beam in $\cQ$. To evaluate the prediction performance, we first apply the following unit step function on each element of the output vector, $f_{\mathrm{step}}(x)= u(x-\delta)$,	where we use $\delta=0.5$ as the threshold. By applying the threshold to $\bt$ as \eqref{thresh}, we obtain $\hat{\bt}$. The predicted UE-side candidate beam set $\hat{\bQ}$ can then be written as
	\begin{align}\label{eq:choose_beam}
		\hat{\bQ} = \left\{ \cQ_{j}| \hat{t}_j =1\right\},
	\end{align}
	where $\hat{t}_j$ is the $j$-th element of $\hat{\bt}$.
	\par
	The metrics adopted to evaluate the performance of the UE-side candidate beam set prediction are the accuracy and the recall. They are defined as follows:
	\begin{equation}
		\mathrm{Acc} = \frac{1}{N_{test}}\sum_{i=1}^{N_{test}} \frac{\left|\bQ^\star_i \cap \hat{\bQ}_i\right|}{\left|\hat{\bQ}_i\right|} \ \ \text{and}  \ \	\mathrm{Recall} = \frac{1}{N_{test}}\sum_{i=1}^{N_{test}} \frac{\left|\bQ^\star_i \cap \hat{\bQ}_i\right|}{\left|\bQ^\star_i \right|},
	\end{equation}
	where $N_{test}$ is the number of data samples in the test dataset. $\bQ^\star_i$ and $\hat{\bQ}_i$ denote the optimal and the predicted UE-side candidate beam set for the $i$-th data sample, respectively.
	
	\begin{figure}[t]
		\centering
		\subfigure[An image from ``Camera 4"]{\includegraphics[width = 0.49\linewidth]{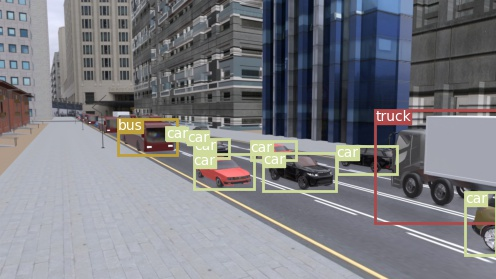}}
		\subfigure[An image from ``Camera 5"]{\includegraphics[width = 0.49\linewidth]{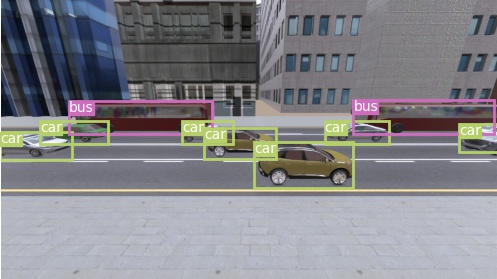}}
		\caption{This figure shows two example images taken by the RIS cameras and illustrates the UE object class and bounding box information which are annotated by the fine-tuned YOLOv3 model. \label{fig:yolo}}
	\end{figure}

	\section{Simulation Results}\label{Simulation}
	In this section, we evaluate the performance of the proposed standalone RIS beam selection.
	First, we present the performance of the YOLOv3 candidate UEs detection.
	Then, we present the accuracy and recall performance of the proposed NN structure in predicting candidate beam sets from the detected candidate UEs. The performance of the proposed NN is compared with two other NN structures that do not reuse the same stack of fully connected layers and the summation operation discussed in \sref{Candidate Beam Set Prediction}.
	Next, we study the amount of data required by the proposed NN in the training process. After that, we demonstrate the effectiveness of the proposed standalone RIS beam selection in terms of the achievable rate. Lastly, we present the proposed RIS beam selection's capability for reducing the beam training overhead.
	\subsection{Can YOLOv3 Detect Candidate UEs?}\label{sim result0}
	Candidate UE detection is the first step of the proposed ML framework for predicting UE-side candidate beam sets. Therefore, the quality of the candidate UE detection is essential for the downstream task and the performance of the framework. Hence, we first demonstrate the performance of the YOLOv3 object detector. In Fig.~\ref{fig:yolo}, we apply the fine-tuned YOLOv3 model on two images from \mbox{``Camera 4"} and \mbox{``Camera 5"}, and let the YOLOv3 model annotate the class and bounding box information on the detected candidate UEs. This figure shows that the YOLOv3 model can accurately detect the candidate UEs in the two cameras and produce high-quality information (class and bounding box) on these UEs.
	\begin{figure}[t]
		\centering
		\includegraphics[width=0.6\linewidth]{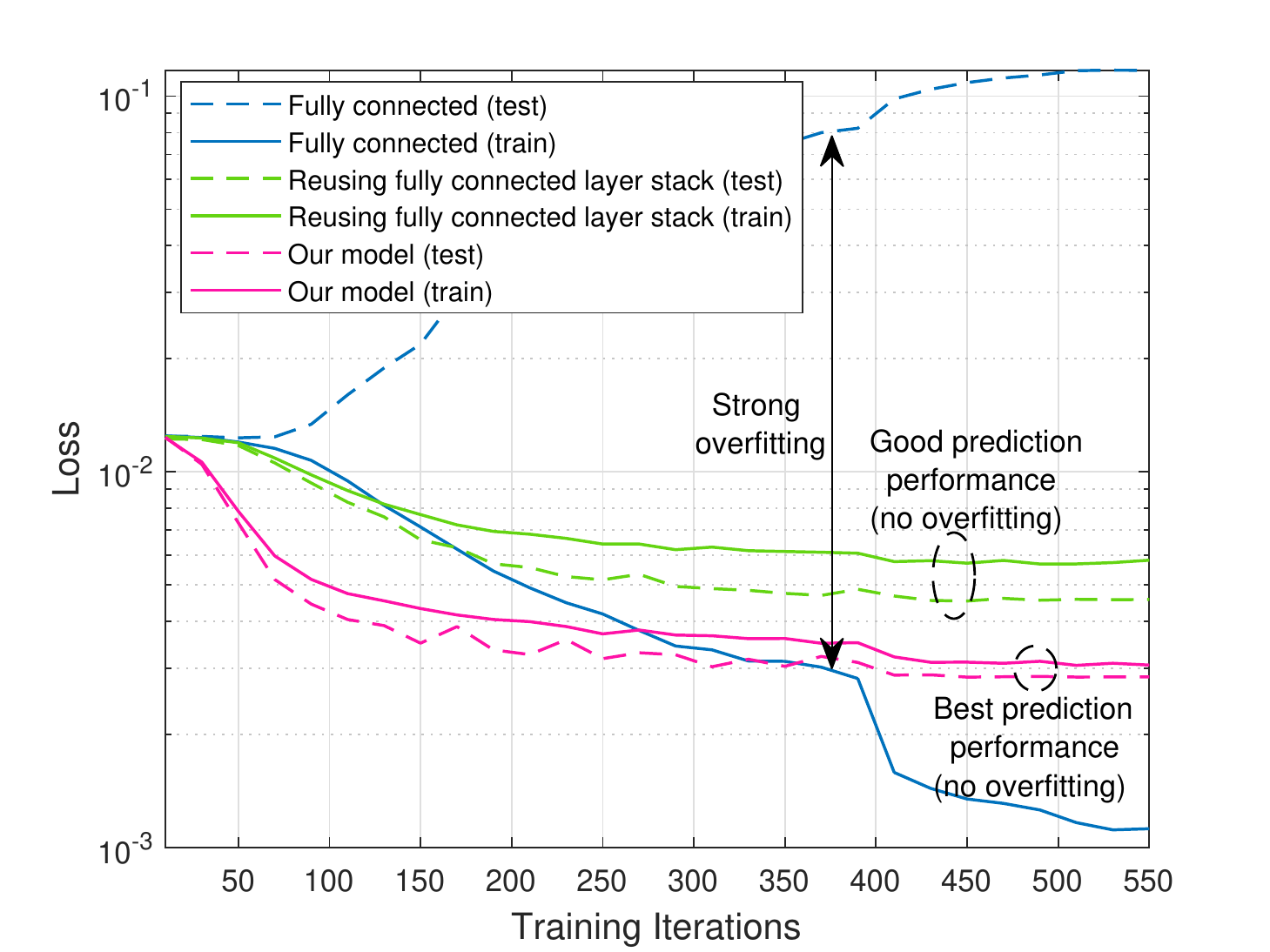}
		\caption{This figure compares the learning curves (train and test) of the proposed neural network structure and two other baseline models when trained on the ``Camera 5" dataset.}
		\label{fig:learning_curve}
	\end{figure}
	
	\subsection{Does the Proposed Neural Network Structure Learn Better?}\label{sim result1}
	In \sref{Candidate Beam Set Prediction}, we mentioned two key features of the proposed NN structure: (i) Reusing the fully connected layer stack on all the UE information vectors, and (ii) combining the information of the detected UEs by the summation operation. These two features are expected to improve the performance of the candidate beam set prediction. To analyze the effectiveness of the proposed NN structure and verify the intuitions used in its design, we study the training process of the NN. Fig.~\ref{fig:learning_curve} presents the learning curves of the proposed NN structure compared with two variants trained on the \mbox{``Camera 5"} dataset. The first variant adopts vanilla fully connected NN. The second variant reuses the same fully connected layer stack on the information vectors from all candidate UEs, but it concatenates the resulting high-level feature vectors instead of applying the summation operation. From the training and the test loss in Fig. \ref{fig:learning_curve}, we see that the vanilla fully connected NN overfits to the training dataset and its loss diverges on the test set. For the second variant, the test loss can converge along with training iterations. This indicates that \textbf{reusing the fully connected layer stack stabilizes the training process}. The proposed NN structure achieves the lowest loss on the test dataset, and the gap between the training and the test losses is the smallest. This implies that \textbf{combining the information from different UEs with the summation operation improves the robustness of the model}. These results highlight the effectiveness of the proposed NN structure over the other architectures. Next, we will further evaluate the proposed NN structure in terms of the accuracy and recall performance.
	
	\begin{figure}[t]
		\centering
		\includegraphics[width=.6\linewidth]{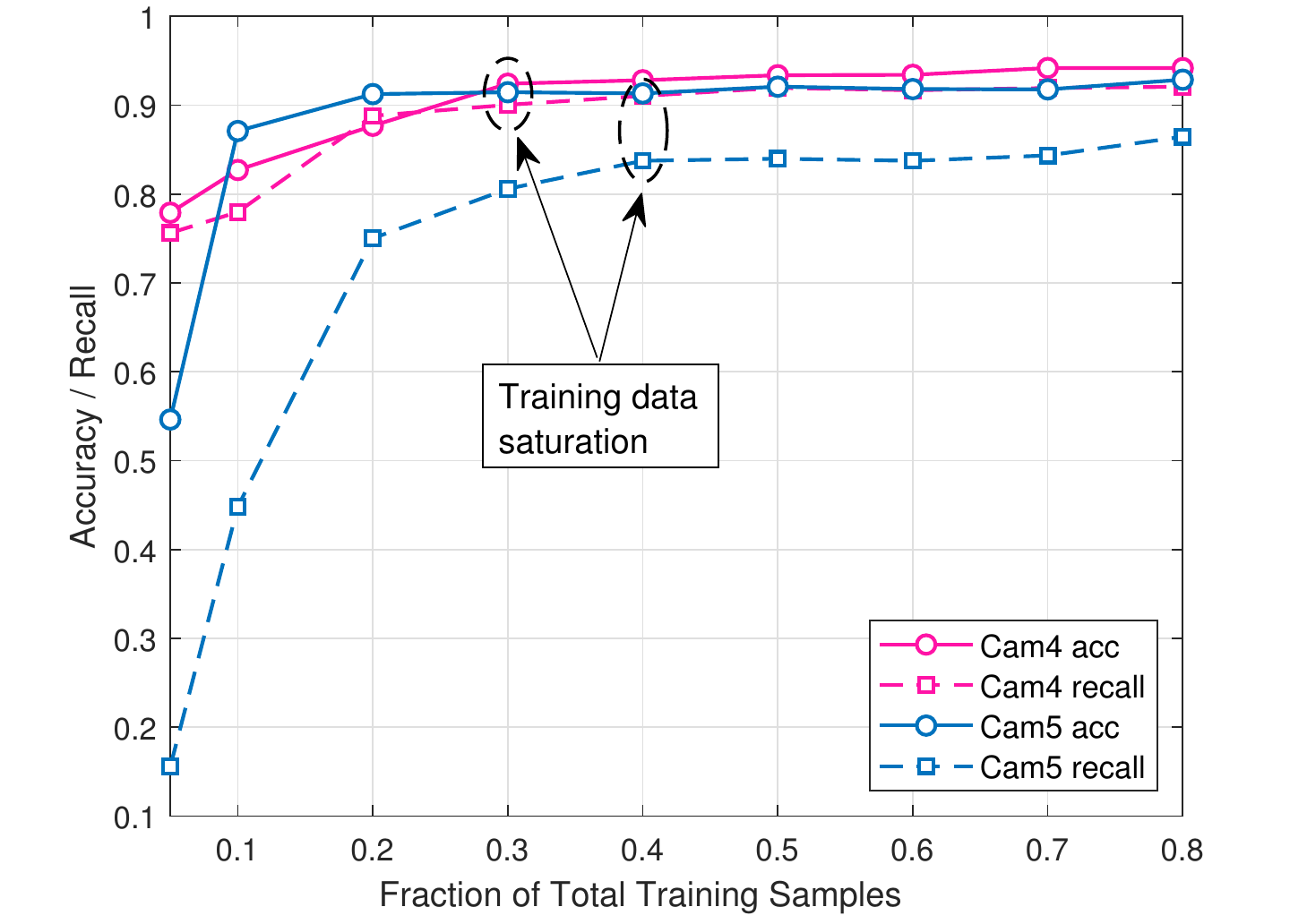}
		\caption{The accuracy and recall performance of the proposed neural network structure when trained on different training dataset sizes (as fractions of the full training set). The figure shows that only 30-40\% of the dataset (which correspond to ~2500-3000 data points) are enough to achieve around 90\% accuracy and recall.}
		\label{fig:acc_recall}
	\end{figure}

	\subsection{Can the Proposed NN Structure Predict the Candidate Beams More Accurately?}\label{sim result2}
	After the YOLOv3 detects the candidate UEs, the proposed NN structure predicts the UE-side candidate beam set. The quality of the predicted candidate beam set directly determines the final transmission rate and the required beam training overhead. Therefore, in Table \ref{tb:finall_performance0}, we present the accuracy and recall performance of the proposed NN structure compared with its two variants. Our proposed NN structure achieves $94.2\%$ and $92.9\%$ on the ``Camera 4'' dataset for accuracy and recall, respectively. On the ``Camera 5'' dataset, the accuracy and recall performance of the proposed NN structure are $92.1\%$ and $86.4\%$. These results highlight that the proposed NN structure can accurately predict the UE-side candidate beam set from the UE information detected by the YOLOv3 model.
	\par
	Comparing the proposed NN structure with the two variants, it can be seen that reusing the fully connected stack offers significant improvements, in terms of accuracy and recall, on both datasets. {For the dataset of ``Camera 5'', the accuracy increases by $\bf{59.5\%}$. On top of that, by reusing the fully connected stack and combining information of candidate UEs with the summation operation, our NN structure results in the highest performance on the test datasets. The recall performance for the ``Camera 5' dataset is improved by $\bf{15.1\%}$.} This again emphasizes that the two key features of the proposed NN structure help stabilize the training process and achieve better performance in the beam set prediction. It is also reinforced that insights on the data can be utilized to improve the performance of ML models.
	\begin{table*}[t]
		\normalsize
		\setlength\tabcolsep{4pt}
		\caption{\label{tb:finall_performance0}Accuracy and recall performance of the ML framework with three different NN structures trained on $80\%$ of the data.}
		\centering
		\begin{small}
			\begin{tabular}[b]{l
					S[table-format=-1.2]
					S[table-format=-1.3]
					S[table-format=-1.2]
					S[table-format=-1.3]
					S[table-format=-1.2]
					S[table-format=-1.3]
					S[table-format=-2.2]
					S[table-format=1.2]
					S[table-format=-2.2]
					S[table-format=1.2]
					S[table-format=-2.2]
					S[table-format=-1.2]}
				\toprule
				\multicolumn{1}{c}{Model}& \multicolumn{2}{c}{Accuracy} & \multicolumn{2}{c}{Recall}\\
				\cmidrule(lr){1-1} \cmidrule(r){2-3} \cmidrule(lr){4-5}
				& {Cam4} & {Cam5} & {Cam4} & {Cam5}\\
				Fully connected& {71.2\%} & {32.6\%} & {70.1\%} & {25.5\%} \\
				Reusing fully connected layer stack&{92.3\%} & {92.1\%} & {87.6\%} &{71.3\%}\\
				Reusing fully connected layer stack + sum operation& {\bf 94.2\%} &{\bf 92.9\%} &{\bf 92.1\%} & {\bf 86.4\%} \\
				\bottomrule
			\end{tabular}
		\end{small}
	\end{table*}
	\subsection{How Much Data is Needed to Train the Beam Prediction NN?}\label{sim result3}
	The size of the training data set is crucial for ML models when deployed in the real world. To that end, in this section, we draw insights on the dataset size required to train the proposed ML framework. Fig. \ref{fig:acc_recall} plots the test accuracy and recall obtained on datasets \mbox{``Camera 4"} and ``Camera 5" versus the fraction of data used to train the proposed UE-side candidate beam prediction NNs. As can be seen from this figure, more training data helps improve the accuracy and recall performance. {The accuracy and recall start to saturate after ${ 30\%}$ and ${ 40\%}$ of data} are used in the training process for \mbox{``Camera 4"} and \mbox{``Camera 5"}, respectively. This corresponds to 2815 data points for \mbox{``Camera 4"} and $2382$ data points for \mbox{``Camera 5"}. Training the proposed NN structure only requires a relatively small dataset since the proposed NN predicts the candidate beam set from the UEs detected by the YOLOv3 model instead of the raw RGB images. Besides, only $500$ samples are used to fine-tune the YOLOv3 model for each dataset. These results show that \textbf{the proposed ML framework is data-efficient in the training process}.
	
	\subsection{How Good is the Achievable Rate?}
	So far, we have demonstrated the intermediate results related to candidate UE detection and candidate beam set prediction. However, the final performance metrics that really matter for the proposed transparent RIS beam selection are the achievable rate and the beam training overhead. Here, we first present the numerical results on the achievable rate.
	\par
	\begin{figure}[t]
		\centering
		\includegraphics[width=0.6\linewidth]{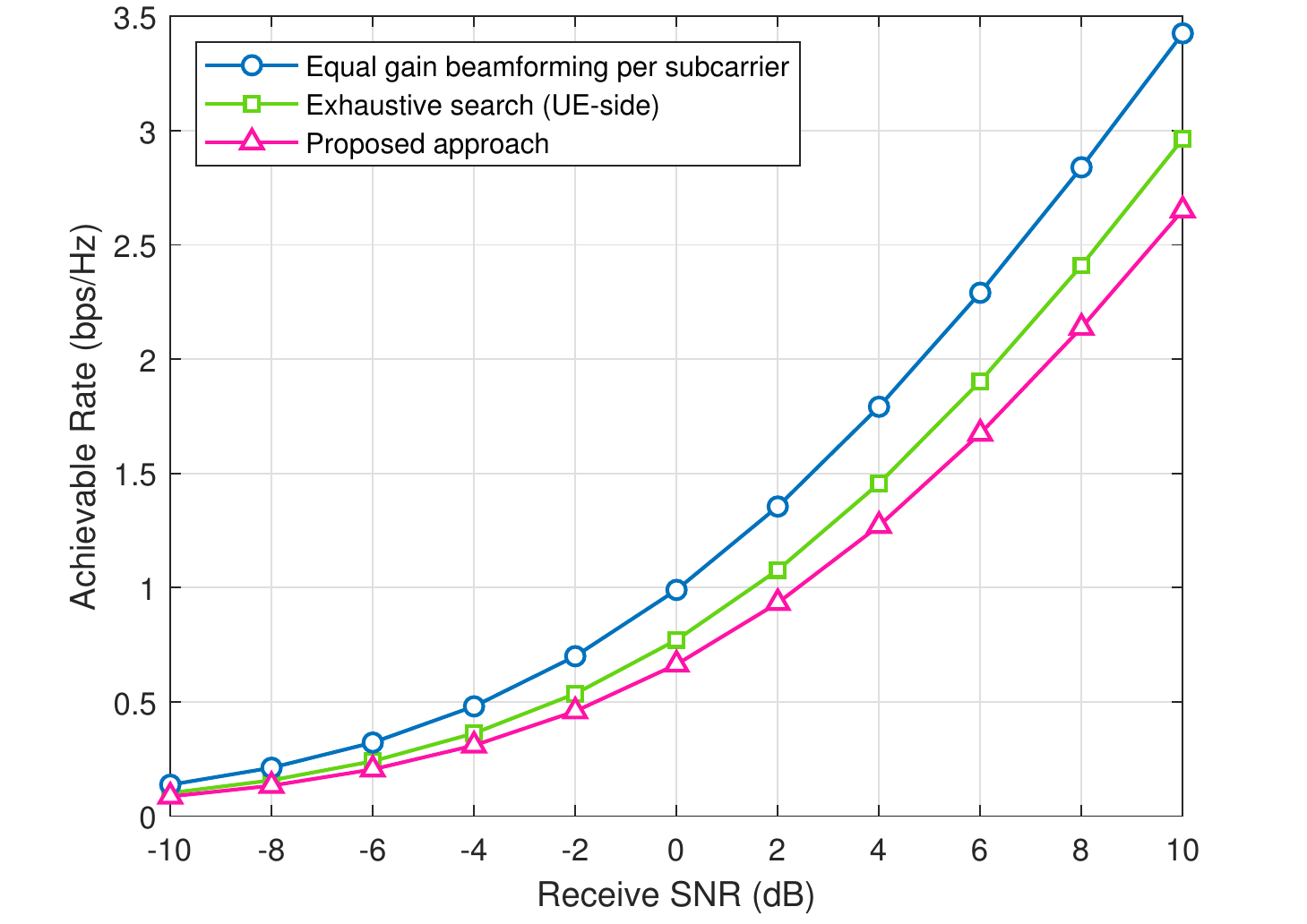}
		\caption{The achievable rate using the beams predicted by the proposed standalone RIS approach compared to (i) the per-subcarrier equal gain beamforming upper bound (which assumes perfect channel knowledge at all RIS elements and all subcarriers) and (ii) the exhaustive search which requires sweeping over all the RIS codebook beams. }
		\label{fig:ach_rate}
	\end{figure}
	Fig. \ref{fig:ach_rate} shows the achievable rate performance of the proposed standalone RIS beamforming compared with two baseline beamforming methods. The first baseline method is the equal gain per-subcarrier beamforming. It uses the conjugate of the channel vectors as the beamforming vectors, which serves as an upper bound of the achievable rate. The second baseline method is the UE-side exhaustive search with a codebook size of $256$. Given a pre-determined BS-side beam, the UE-side exhaustive search tries all UE-side beams in the UE-side codebook to obtain the highest achievable rate. Note that both the two baseline beamforming methods are not practical. The equal gain combine applies different beamforming on different subcarriers, which violates the implementation constraint of the RIS. The exhaustive search requires either the channel information on both the BS-side and the UE-side, or the beam training (sweeping) process. In the UE-side exhaustive search and our approach, the optimal BS-side beam $\bp^{\star}$ in \eqref{eq:best_beam_easy_a} is used. As can be seen in Fig. \ref{fig:ach_rate}, the achievable rate increases as the SNR becomes higher for all three methods. On top of that, we see that \textbf{our proposed approach obtains high performance in the SNR range from $\bf -10$ to $\bf 10$ dB compared with these two  upper bounds. For example, at $0$ dB receive SNR, It achieves $\bf 86.1\%$ of the exhaustive search data rate}.
	
	\subsection{How Much Beam Training Overhead is Required?}
\begin{figure}[t]
		\centering
		\includegraphics[width=0.6\linewidth]{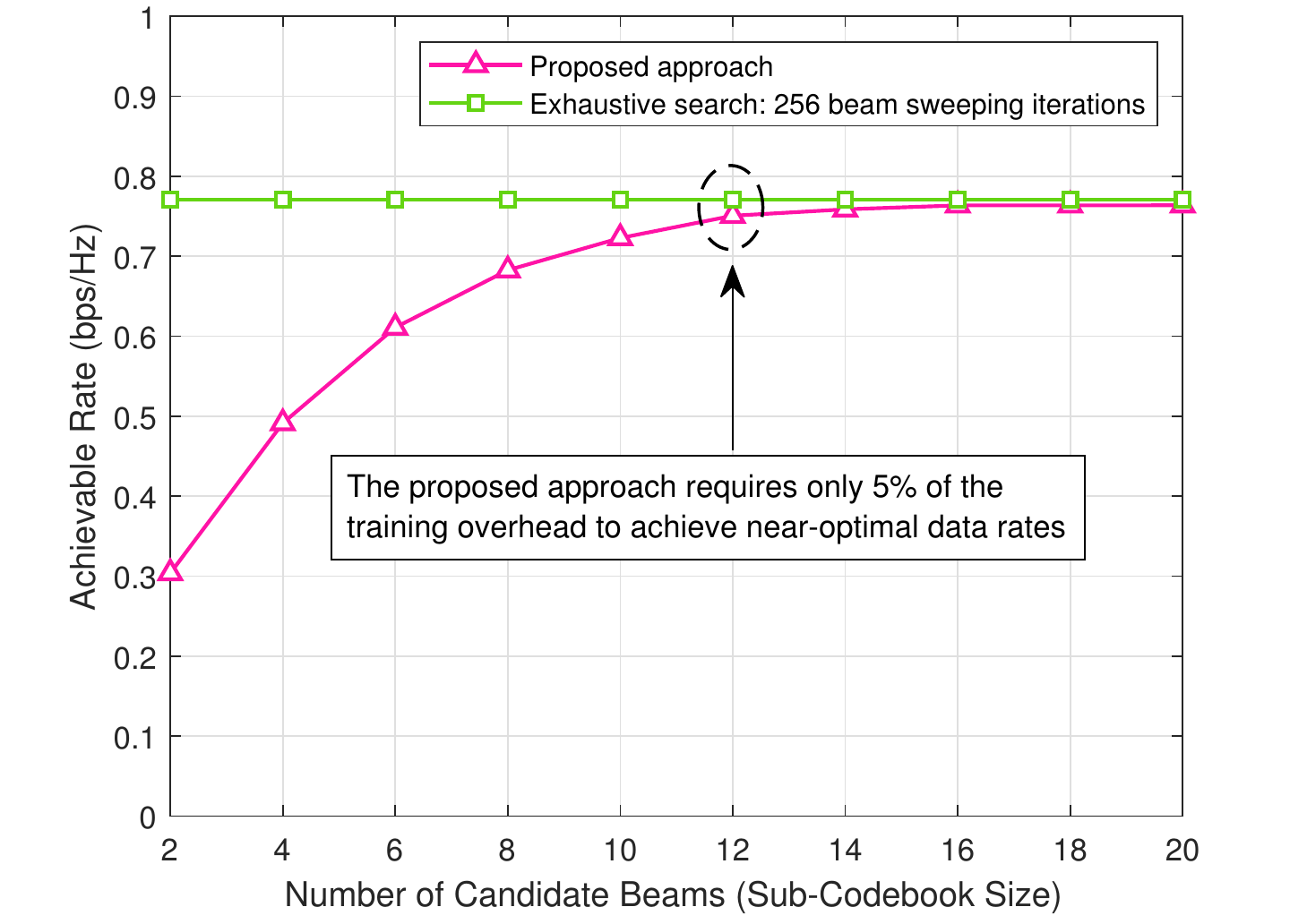}
		\caption{The achievable rate of the proposed standalone RIS approach with different sizes of the UE-side candidate beam set. The upper bound achievable rate is obtained by the exhaustive search over all the beams in the RIS codebook.}
		\label{fig:ach_rate_beam_set_size}
	\end{figure}

	Apart from the high achievable rate, another goal of this paper is to reduce the beam training overhead of the standalone RIS. Thus, in Fig. \ref{fig:ach_rate_beam_set_size}, we investigate the effect of the size of the UE-side candidate beam set on the achievable rate at $0$ dB receive SNR. In the previous simulations we apply the step function in \sref{subsec:metrics} to the output vector of the NN, and construct the candidate beam set as shown by \eqref{eq:choose_beam}. Here, for each scene in the datasets, the candidate beam set consists of the $k$ beams corresponding to the top-$k$ highest value in $\bf t$, the output vector of the NN. As shown in Fig. \ref{fig:ach_rate_beam_set_size}, when the candidate beam set size increases, the achievable rate performance of our proposed standalone RIS beam selection approaches the performance of the UE-side exhaustive search. \textbf{When only $\bf 12$ out of $\bf 256$ beams are used in the UE-side candidate beam set, the proposed approach can achieve $\bf 97.4\%$ of the exhaustive search data rate}. Note that the UE-side exhaustive search provides the {optimal achievable rate} when a BS-side beam is given. Therefore, approaching this upper bound indicates that the proposed vision aided standalone RIS can efficiently reduce the beam training overhead with little negative effect on the achievable rate of the RIS-aided communication system.
	
	\section{Discussion and Future Work}\label{Future Work}
	In \sref{Vision-Aided Transparent RIS for 5G 3GPP}, we elaborated on the proposed transparent  3GPP 5G beamforming operation of the vision-aided RIS system. In this section, we provide some insights on how to extend the proposed  3GPP 5G transparent RIS operation to more dynamic and complex scenarios, and discuss the enabling features.
	\subsection{Beam Tracking}
	The ability of the RIS to perform efficient beam tracking is essential for supporting mobile UEs in dynamic environments. The 3GPP 5G beam refinement process in \cite{3GPP802} includes the following: (i) The BS performs beam sweeping within a subset of transmit beams while the UE is maintaining a receive beam, and (ii) the UE  reports the beam measurement results and the selected beam(s) to the BS. This beam refinement process, however, cannot be directly employed by the standalone RIS operation. The beam refinement requires signaling between the RIS and the BS/UE because the RIS lacks knowledge of the content of some messages exchanged between the BS and UE, \textit{e.g.}, the ``Message 1" and ``Message 2". As discussed in \sref{Vision-Aided Transparent RIS for 5G 3GPP}, however, the RIS's sensing capability can be utilized to achieve transparent RIS beam tracking. With the sensing capability, the RIS can obtain rich information about the UE position, its mobility pattern, and the environment layout, from which the RIS can infer its future UE-side beam from the current/previous beam sequence.
	\subsection{Beam Tracking Stopping Criteria}
	The RIS should stop the beam tracking when it detects that the communication between the UE and the BS ends. As discussed in \sref{Vision-Aided Transparent RIS for 5G 3GPP}, the RIS beam tracking stopping criteria include: (i) There is a blockage between the RIS and the BS/UE, (ii) the BS has switched to other UEs, and (iii) the communication session is terminated. Note, however, that since the standalone RIS does not communicate with the BS or the UE, it can not directly know that the session has ended and it also can not access the configuration of the wireless resource allocation. Therefore, the RIS  needs to infer and keep track of the wireless resource associated with the UE. Having a spectrum sensing capability at the RIS may be one step towards this objective. It remains, however, an interesting and open research problem. Apart from the spectrum sensing approach, other sensing modalities at the RIS can also be leveraged to detect the beam tracking criteria. For instance, the sensors (such as camera, radar, LiDAR, etc.) can detect and proactively predict the potential blockages between the BS/UE and the RIS and assist the stopping criteria detection for the standalone RIS beam tracking.
	\subsection{Multi-User Scenario}
	In the 3GPP 5G, one BS can serve multiple UEs by allocating different time, frequency, and beam resources to different UEs. The proposed transparent RIS operation in \sref{Vision-Aided Transparent RIS for 5G 3GPP} focuses on the single-user scenario. However, it is important to extend that to the multi-user scenarios. When the UEs are configured to the same time resource, the RIS can only apply the same reflecting beam at the same time instance due to the time-domain implementation limitation. Therefore, \textit{one} RIS reflecting beam may be designed to simultaneously serve multiple UEs. Alternatively, the RIS can be divided into multiple sub-arrays to serve different UEs with different reflecting beams. The multi-user transparent RIS beam selection is still an open problem that requires future research consideration.
	\section{Conclusion}\label{Conclusion}
	In this paper, we investigated the feasibility of enabling  3GPP 5G transparent RIS operation using sensing-based perception. Utilizing the sensing capability at the RIS, we proposed a standalone RIS beam selection operation that does not need any dedicated signaling with the user or the basestation, and is compatible with the  3GPP 5G initial access process. To guide the RIS beam selection in the proposed standalone RIS operation, we developed a machine learning framework and a neural network architecture that leverage the visual data captured by cameras installed at the RIS to predict the candidate set of beams. To evaluate the developed sensing-aided machine learning solution, we conducted extensive simulations based on a high-fidelity synthetic dataset gathering co-existing wireless and visual data.  When benchmarked against other neural network architectures, the proposed one shows clear advantage in both learning stability and prediction accuracy. In particular, the simulation results demonstrated that the proposed machine learning framework can accurately predict the user-side candidate beam set with high accuracy and recall performance. Further, these results showed that the proposed standalone RIS system can achieve near-optimal spectral efficiency with significantly reduced beam training overhead. This highlights the potential of leveraging sensing based perception to develop  3GPP 5G transparent RIS operation. For the future work, it is interesting to explore how this work can be extended to account for beam tracking and maintenance (after the link is established) and for multi-user and multi-basestation settings. 
	\appendices
	\section{}\label{Appendix}
	When the BS-side channel and UE-side channel only contain the LoS path, all the subcarriers have the same channels, \textit{i.e.}, $\bh_{\mathrm{T},k} = \bh_{\mathrm{T}}$, $\forall k=1,\hdots,K$, and $\bh_{\mathrm{R},k} = \bh_{\mathrm{R}}$, $\forall k=1,\hdots,K$. Without loss of generality, let us assume
	\begin{equation}
		\bh_{\mathrm{T}} = \left[ a_1e^{j\phi_1}, \hdots, a_Me^{j\phi_M} \right]^T,
		\bh_{\mathrm{R}} = \left[ b_1e^{j\omega_1}, \hdots, b_Me^{j\omega_M} \right]^T, 
		{\ba} = \left[ c_1e^{j\beta_1}, \hdots, c_Me^{j\beta_M} \right]^T,
	\end{equation}
	where $a_m,b_m,c_m \geq 0$, and $\phi_m, \omega_m \in[-\pi,\pi)$, \mbox{$\forall m=1,\hdots, M$}.
	Since ${\bm \psi}^\star, {\bp}^\star, {\bq}^\star \in \cO$, they can be written as
	\begin{equation}
		{\bm \psi}^\star = \left[ e^{j\lambda_1}, \hdots, e^{j\lambda_M} \right]^T, 
		{\bp}^\star = \left[ e^{j\lambda_{p,1}}, \hdots, e^{j\lambda_{p,M}} \right]^T, 
		{\bq}^\star = \left[ e^{j\lambda_{q,1}}, \hdots, e^{j\lambda_{q,M}} \right]^T,
	\end{equation}
	where $\lambda_m, \lambda_{p,m}, \lambda_{q,m},\beta_m \in[-\pi,\pi)$, $\forall m=1,\hdots, M$. With $\bh_{\mathrm{T},k} = \bh_{\mathrm{T}}$ and $\cP = \cO$, \eqref{eq:best_beam_easy_a} can be re-written as
	\begin{align}
		{\bp}^\star &= \underset{\bp \in \cO }{\arg\max}\frac{1}{K}\sum_{k=1}^{K}\left| \left(\bh_{{\rm T}}\odot \bp\right)^H \ba^* \right|^2 = \underset{\bp \in \cO }{\arg\max}\left| \left(\bh_{{\rm T}}\odot \bp\right)^H \ba^* \right|^2 \nonumber\\
		&= \underset{\phi_m, \beta_m, \lambda_{p,m}}{\arg\max} \left| \sum_{m=1}^{M} a_m c_m e^{-j(\phi_m+\beta_m+\lambda_{p,m})}\right|^2\\
		\Leftrightarrow& \ \  \phi_{m_1} + \beta_{m_1} +\lambda_{p,{m_1}} = \phi_{m_2} + \beta_{m_2} +\lambda_{p,{m_2}}, \forall m_1, m_2. \label{eq:a1}
	\end{align}
	Similarly, \eqref{eq:best_beam_easy_b} can be re-written as
	\begin{align}
		{\bq}^\star &= \underset{\bq \in \cO}{\arg\max}\,\left| \left(\bh_{{\rm R}}\odot \bq \right)^H{\ba}\right|^2 = \underset{\phi_m, \beta_m, \lambda_{q,m}}{\arg\max} \left| \sum_{m=1}^{M} b_m c_m e^{-j(\phi_m-\beta_m+\lambda_{q,m})}\right|^2\\
		\Leftrightarrow& \ \ \omega_{m_1} - \beta_{m_1} +\lambda_{q,{m_1}} = \omega_{m_2} - \beta_{m_2} +\lambda_{q,{m_2}}, \forall m_1, m_2.\label{eq:a2}
	\end{align}
	Adding up \eqref{eq:a1} and \eqref{eq:a2}, we can derive
	\begin{equation}\label{eq:a3}
		\phi_{m_1}+ \omega_{m_1} +\lambda_{p,{m_1}}+\lambda_{q,{m_1}} 
		= \phi_{m_2} + \omega_{m_2} +\lambda_{p,{m_2}}+\lambda_{q,{m_2}}, \forall m_1, m_2.
	\end{equation}
	Similarly, \eqref{eq:best_beam} can be re-written as
	\begin{equation}\label{eq:a4}
		{\bm \psi}^\star = \underset{{{\bm \psi}\in \cO}}{\arg\max} \, \left| \left(\bh_{{\rm R}}\odot \bh_{{\rm T}}\right)^T{\bm \psi}\right|^2 
		\Leftrightarrow \phi_{m_1}+ \omega_{m_1} +\lambda_{m_1} = \phi_{m_2} +\omega_{m_2} +\lambda_{{m_2}}, \forall m_1, m_2.
	\end{equation}
	Since ${\bp}^\star\odot{\bq}^\star = \left[ e^{j(\lambda_{p,1}+\lambda_{q,1})}, \hdots, e^{j(\lambda_{p,M}+\lambda_{q,M})} \right]^T$, it can be seen from \eqref{eq:a3} that ${\bp}^\star\odot{\bq}^\star$ satisfies \eqref{eq:a4}, therefore, ${\bp}^\star\odot{\bq}^\star$ is an optimal solution of \eqref{eq:best_beam}.

	\linespread{1.2}

\end{document}